\documentclass[a4paper,11pt]{article}
\pdfoutput=1
\usepackage{jheppub}
\usepackage{amsmath}
\usepackage{hyperref}
\usepackage{color,soul}
\allowdisplaybreaks
\def\bea#1\eea{\begin{align}#1\end{align}}
\newcommand{\nnu}{\nonumber\\}
\newcommand{\kt}{anti-k$_{\rm T}$}
\newcommand{\bef}{\begin{figure}[hbt]\centering}
\newcommand{\eef}{\end{figure}}
\newcommand{\sla}[1]{{#1}\!\!\!\slash}
\newcommand{\GG}{{\cal G}}
\newcommand{\R}{{\mathcal R}}
\newcommand{\f}{\frac}
\def \be  {\begin{equation}}
\def \ee  {\end{equation}}
\def \ba  {\begin{eqnarray}}
\def \ea  {\end{eqnarray}}
\newcommand\as{\alpha_s}

\title{Jet substructure using semi-inclusive jet functions in SCET}

\author[a,b]{Zhong-Bo Kang,}
\author[a]{Felix Ringer}
\author[a]{and Ivan Vitev}

\affiliation[a]{Theoretical Division, 
                     Los Alamos National Laboratory,
                     Los Alamos, NM 87545, USA}
\affiliation[b]{Department of Physics and Astronomy, 
                     University of California,
                     Los Angeles, CA 90095, USA}                     
                                         
\emailAdd{zkang@lanl.gov}
\emailAdd{f.ringer@lanl.gov}
\emailAdd{ivitev@lanl.gov}

\abstract{We propose a new method to evaluate jet substructure observables in inclusive jet measurements, based upon semi-inclusive jet functions in the framework of Soft Collinear Effective Theory (SCET). As a first example, we consider the jet fragmentation function, where a hadron $h$ is identified inside a fully reconstructed jet. We introduce a new semi-inclusive fragmenting jet function ${\mathcal G}^h_i(z= \omega_J/\omega,z_h=\omega_h/\omega_J,\omega_J, R,\mu)$, which depends on the jet radius $R$ and the large light-cone momenta of the parton `$i$' initiating the jet ($\omega$), the jet ($\omega_J$), and the hadron $h$ ($\omega_h$). The jet fragmentation function can then be expressed as a semi-inclusive observable, in the spirit of  actual experimental measurements, rather than as an exclusive one. We demonstrate the consistency of the effective field theory treatment and standard perturbative QCD calculations of this observable at next-to-leading order (NLO). The renormalization group (RG) equation for the semi-inclusive fragmenting jet function ${\mathcal G}_i^h(z,z_h, \omega_J, R,\mu)$ are also derived and shown to follow exactly the usual timelike DGLAP evolution equations for fragmentation functions. The newly obtained RG equations can be used to perform the resummation of single logarithms of the jet radius parameter $R$ up to next-to-leading logarithmic (NLL$_R$) accuracy. In combination with the fixed NLO calculation, we obtain NLO+NLL$_R$ results for the hadron distribution inside the jet. We present numerical results for $pp\to(\mathrm{jet}\,h)X$ in the new framework, and find excellent agreement with existing LHC experimental  data.}

\begin{document}
\maketitle

\section{Introduction}
\label{sec:intro}

At present day hadron colliders, collimated jets of hadrons are abundantly produced and, hence, it is not surprising that a large fraction of the observables currently investigated  at the Large Hadron Collider (LHC) involve jets. Experiments at the LHC measure both the inclusive jet production cross sections, as well as a variety of jet substructure observables. In recent years, jet substructure measurements have emerged as an ideal testing ground to study the fundamental properties of Quantum Chromodynamics (QCD). In addition, they provide promising new observables for searches of new physics beyond the Standard Model~\cite{Altheimer:2013yza,Adams:2015hiv}. It is, therefore, of utmost importance to continuously improve our theoretical understanding of the QCD dynamics that govern both the total production cross section of jets~\cite{Ellis:1990ek,Aversa:1988vb,Jager:2004jh,Mukherjee:2012uz,Currie:2013dwa,deFlorian:2013qia,Dasgupta:2014yra,Dasgupta:2016bnd,Kang:2016mcy}, as well as jet substructure~\cite{Ellis:1992qq,Seymour:1997kj,Ellis:2010rwa,Procura:2009vm,Liu:2010ng,Jain:2011xz,Jain:2011iu,Procura:2011aq,Li:2011hy,Li:2012bw,Chien:2012ur,Chien:2014nsa,Hornig:2016ahz,Bauer:2013bza,Cacciari:2012mu,Ritzmann:2014mka,Baumgart:2014upa,Chien:2015ctp,Bain:2016clc,Arleo:2013tya,Kaufmann:2015hma,Kaufmann:2016nux,Kolodrubetz:2016dzb}.

In this work we propose a new formalism to study jet substructure observables in Soft Collinear Effective Theory (SCET)~\cite{Bauer:2000ew,Bauer:2000yr,Bauer:2001ct,Bauer:2001yt,Bauer:2002nz}, based upon  semi-inclusive jet functions recently introduced to describe inclusive jet spectra~\cite{Kang:2016mcy}. With the help of these semi-inclusive jet functions we are able to express jet substructure observables related to inclusive jet measurements $pp\to\mathrm{jet}X$, where the sum over all particles in the final state $X$ besides the observed jet is performed. At present, many jet substructure observables are calculated for exclusive processes $pp\to n\,\mathrm{jets}$ in SCET. Here, ``exclusive'' means that a certain number of signal jets is identified but one vetoes additional jets. To achieve this, an upper energy cut $\Lambda$ for the total energy outside the observed $n$-jets in the final state can be imposed. There are important exclusive-type jet measurements at both the Tevatron and the LHC, e.g., exclusive jet production associated with vector bosons~\cite{Abazov:2008ez,Group:2010aa,Chatrchyan:2011ne,Aad:2013ysa}. However, lots of jet cross section and jet substructure measurements, e.g.
jet mass, jet shapes, and jet fragmentation functions,  are usually performed for inclusive jet production~\cite{CMS:2011ab,Aad:2011fc,Chatrchyan:2012mec,ATLAS:2012am,ALICE:2014dla,ATLAS:2015mla,Aad:2011sc,Aad:2011td,Chatrchyan:2012gw,Aad:2014wha,Chatrchyan:2014ava,Abelev:2013kqa}. For such observables,  one should ideally use a factorization formalism developed specifically for inclusive jet production, rather than rely on a framework formulated for  exclusive jets. Besides the fact that this will reflect more accurately the nature of the experiments at the LHC, it will also help eliminate potentially large power corrections of the form ${\cal O}(\Lambda/Q)$, where $Q$ denotes the hard scale of the process.

As a first example of a jet substructure observable, we consider the jet fragmentation function (JFF) expressed as a semi-inclusive measurement in SCET. The JFF describes the longitudinal momentum distribution of hadrons inside a fully reconstructed jet. The exact definition will be given below. The JFF probes the parton-to-hadron fragmentation at a differential level and it can give novel constraints for fragmentation functions. In addition, one may gain new insights into spin dependent phenomena~\cite{Yuan:2007nd,DAlesio:2010am,DAlesio:2011mc,Aschenauer:2013woa,Aschenauer:2015eha}. In heavy-ion collisions the JFF plays an important role where one may study the modification of jets when traversing the hot and dense QCD medium, the quark-gluon plasma~\cite{Borghini:2005em,Casalderrey-Solana:2015vaa,medium_jet_frag}, and strongly complements the modification
of other jet cross section and jet substructure observables~\cite{He:2011pd,Chien:2015hda,Chang:2016gjp,Senzel:2016qau}.  Experimentally, the JFF was first measured in $p\bar p$ collisions at the Tevatron~\cite{Abe:1990ii}, and more recently in $pp$ collisions at the LHC~\cite{Aad:2011sc,Aad:2011td,Aad:2014wha,Chatrchyan:2012gw,Chatrchyan:2014ava}. Pioneering work for the theoretical description of the JFF was performed in~\cite{Procura:2009vm,Liu:2010ng,Jain:2011xz,Jain:2011iu,Procura:2011aq} using the framework of SCET. Further investigation and extensions were presented in~\cite{Bauer:2013bza,Cacciari:2012mu,Arleo:2013tya,Ritzmann:2014mka,Baumgart:2014upa,Kaufmann:2015hma,Chien:2015ctp,Bain:2016clc,Kaufmann:2016nux} using both SCET and  standard perturbative QCD methods. In particular, the JFF in $pp$ collisions was calculated for inclusive $pp\to(\mathrm{jet}h)X$ processes to fixed next-to-leading order (NLO) in~\cite{Arleo:2013tya,Kaufmann:2015hma,Kaufmann:2016nux}. In~\cite{Chien:2015ctp}, the JFF in $pp$ collisions was addressed in the context of SCET. While the effective field theory treatment in~\cite{Chien:2015ctp} allowed for the resummation of potentially large logarithms in the jet parameter $R$, which is not achieved by fixed NLO calculations, it was written in terms of exclusive jet functions. In contrast, in this work we present a calculation of the JFF $pp\to(\mathrm{jet}h)X$ in SCET written as a semi-inclusive cross section ratio, which is consistent with the fixed NLO results of~\cite{Arleo:2013tya,Kaufmann:2015hma,Kaufmann:2016nux}. In addition, owing to the effective field theory treatment, we are able to go beyond the current state-of-the-art fixed order calculation  by resumming potentially large logarithms of the jet radius parameter $R$  through renormalization group (RG) equations.

In order to write the JFF in SCET for inclusive jet measurements, we introduce a new type of jet function -- the semi-inclusive fragmenting jet function~\cite{KRV} (FJF) $\GG_i^h(z,z_h,\omega_J,R,\mu)$. Here, $\mu$ is the renormalization scale and $R$ is the jet radius parameter. We further define the following three large light-cone momentum components $\omega_J,\,\omega,\,\omega_h$ which correspond to the jet, the parton $i$ initiating the jet and the hadron observed inside the jet respectively. The variables $z,z_h$ are given by the ratios $z = \omega_J/\omega$ and $z_h=\omega_h/\omega_J$. We derive the RG equations for the semi-inclusive FJF, which take the form of standard timelike DGLAP equations that also govern the evolution of fragmentation functions. By solving the DGLAP equations, we are able to resum single logarithms of the jet radius parameter $\alpha_s^n \ln^n R$ up to next-to-leading logarithmic (NLL$_R$) accuracy and combine it  with the fixed order results to obtain NLO+NLL$_R$. This needs to be contrasted to the exclusive limit of the JFF, where the dependence on the jet radius parameter is double logarithmic, i.e. $\alpha_s^n \ln^{2n} R$, and resummation proceeds through a multiplicative RG equation~\cite{Chien:2015ctp}. In~\cite{Kang:2016mcy}, an analogous new kind of jet function $J_i(z,\omega_J,\mu)$ was introduced in order to describe single inclusive jet production in SCET $pp\to\mathrm{jet}X$. For related work on single-inclusive jet production, see also~\cite{Dasgupta:2014yra,Dasgupta:2016bnd, Dai:2016hzf}. Understanding the underlying dynamics of small-$R$ jets is particularly relevant for jet substructure studies in heavy-ion collisions, where the experiments typically choose a very small jet parameter in order to minimize the contribution of background radiation. For example, in~\cite{Abelev:2013kqa,Chatrchyan:2014ava,Aad:2014wha}, the jet parameter is chosen as $R=0.2$ and $R=0.3$.

The remainder of this paper is organized as follows. In section~\ref{sec:two}, we provide the definition of the new semi-inclusive FJF and  give details of its evaluation  to first order in the strong coupling for both the cone and  \kt~algorithms. In addition, we derive the DGLAP type RG equations and we discuss their solution in Mellin moment space. In section~\ref{sec:three}, we present numerical calculations using our new framework. We first compare to the currently available data from the LHC, and, we then present comparisons to the fixed NLO results for the JFF. We conclude our paper in section~\ref{sec:four}.

\section{The semi-inclusive fragmenting jet function \label{sec:two}}
In this section, we introduce the definition of the semi-inclusive fragmenting jet function in SCET, perform its calculation to NLO, and, finally, derive and solve its RG evolution equation. 

\subsection{Definition}
The semi-inclusive fragmenting quark and gluon jet functions can be constructed from the corresponding gauge invariant quark and gluon fields in SCET, which are given by~\cite{Bauer:2000ew,Bauer:2000yr,Bauer:2001ct,Bauer:2001yt,Bauer:2002nz}
\bea
\chi_{n} =  W_n^\dagger \xi_n,
\qquad
{\mathcal B}_{n\perp}^\mu =  \frac{1}{g}\left[W_n^\dagger iD_{n\perp}^\mu W_n\right].
\label{eq:collinear-fields}
\eea
Here, $n^\mu$ is a light-cone vector with its spatial component along the jet axis. It is convenient to introduce another conjugate light-cone vector $\bar n^\mu$, such that $n^2 = \bar n^2 = 0$ and $n\cdot \bar n = 2$. In Eq.~\eqref{eq:collinear-fields}, the covariant derivative is $iD_{n\perp}^\mu = {\mathcal P}_{n\perp}^\mu + gA_{n\perp}^\mu$, with ${\mathcal P}^\mu$ the label momentum operator. On the other hand, $W_n$ is the Wilson line of collinear gluons,
\bea
W_n(x) = \sum_{\rm perms} \exp\left[-g\frac{1}{\bar n\cdot {\mathcal P}} \bar n\cdot A_n(x)\right].
\eea

With these collinear quark and gluon fields at hand, the semi-inclusive FJFs for quark and gluon jets are defined as 
\bea
\GG_q^h(z, z_h, \omega_J, \mu) =& \frac{z}{2N_c} \delta\left(z_h - \frac{\omega_h}{\omega_J}\right)
{\rm Tr} \left[\frac{\sla{\bar n}}{2}
\langle 0| \delta\left(\omega - \bar n\cdot {\mathcal P} \right) \chi_n(0)  |(Jh)X\rangle \langle (Jh)X|\bar \chi_n(0) |0\rangle \right],
\\
\GG_g^h(z, z_h, \omega_J, \mu) =& - \frac{z\,\omega}{(d-2)(N_c^2-1)} \delta\left(z_h - \frac{\omega_h}{\omega_J}\right)
\langle 0|  \delta\left(\omega - \bar n\cdot {\mathcal P} \right) {\mathcal B}_{n\perp \mu}(0) 
 |(Jh)X\rangle 
 \nnu
&\hspace{50mm} \times \langle (Jh)X|{\mathcal B}_{n\perp}^\mu(0)  |0\rangle,
\eea
where $(d-2)$ is the number of polarizations for gluons in $d$ space-time dimensions. Note that we only consider massless quark flavors. The state $|(Jh)X\rangle$ represents the final-state unobserved particles $X$ and the observed jet $J$ with an identified hadron $h$ inside, denoted collectively by $(Jh)$. On the other hand, $\omega$ is the quark or gluon energy initiating the jet, while $\omega_J$ and $\omega_h$ are the energy of the jet and that of the identified hadron inside the jet, respectively. The energy fractions $z$ and $z_h$ are defined as follows
\be
z=\f{\omega_J}{\omega}\, ,\quad z_h=\f{\omega_h}{\omega_J} \, .
\ee
Note that the variable $z$ also appears in the closely related calculation of the semi-inclusive jet function presented in~\cite{Kang:2016mcy}. In addition, we would like to point out that the semi-inclusive fragmenting jet function can also depend on the jet radius $R$, i.e. we have in general $\GG_i(z,z_h,\omega_J,R,\mu)$. However, in the remainder of this paper we leave this dependence implicit to shorten our notation.

\subsection{NLO calculation}
Since the semi-inclusive FJFs $\GG_{i}^h(z, z_h, \omega_J, \mu)$ describe the distribution of hadrons inside the jet, and, thus, contain hadronization/non-perturbative information, they are not directly calculable in perturbation theory. In this respect, they are different from the purely perturbative semi-inclusive jet functions introduced in \cite{Kang:2016mcy}. Nevertheless, following the standard perturbative QCD methodology, one can evaluate the partonic fragmenting jet functions to obtain their renormalization properties. In other words, we replace the hadron $h$ by a parton $j$, and compute $\GG_{i}^j(z, z_h, \omega_J, \mu)$ as a perturbative expansion in terms of the strong coupling constant $\alpha_s$. 

We are now going to outline the calculation of the semi-inclusive FJF for quark and gluon initiated jets $\GG_{q,g}^h(z,z_h,\omega_J,\mu)$. Although the results depend on the jet algorithm, to facilitate our presentation, we mostly focus on the calculation for the \kt~algorithm. We only list results for the cone algorithm at the end, see Eq.~(\ref{eq:nloresult}), and we point out important differences along the way. At leading order the results only involve two delta functions
\bea\label{eq:GLO}
{\cal G}_q^{q,(0)}(z, z_h,\omega_J) = {\cal G}_g^{g,(0)}(z, z_h,\omega_J)  = \delta(1-z)\delta(1-z_h) \, .
\eea
Note that $z=1$ corresponds to the case where the total energy of the initiating parton is transferred to the jet. On the other hand, $z_h=1$ corresponds to the case, where the fragmenting parton inside the jet carries the total jet energy. Both quantities are unity at leading-order but they will have a more complicated functional form at NLO and beyond,  allowing generally $z,z_h<1$.
\bef
\includegraphics[width=\textwidth]{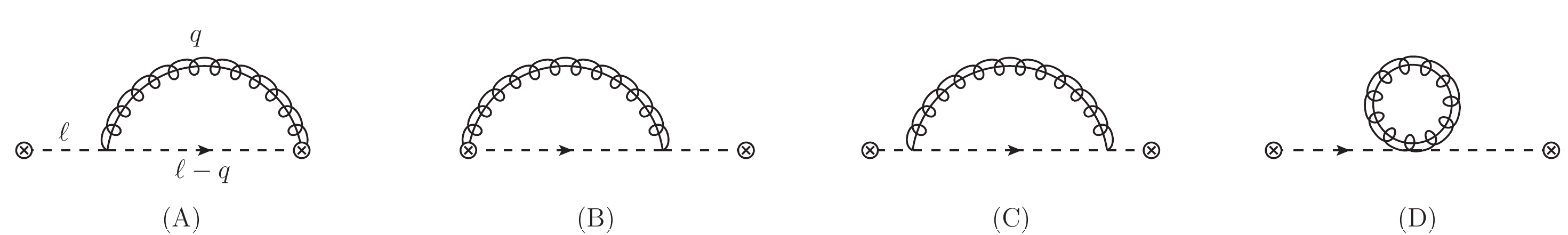}
\caption{Feynman diagrams that contribute to the semi-inclusive quark fragmenting jet function. The quark initiating the jet has momentum $\ell = (\ell^+, \ell^- = \omega, 0_\perp)$, with $\omega = \omega_J/z=\omega_h/(zz_h)$ and $\omega_J$, $\omega_h$ are the jet and hadron energies respectively. Note that the dashed (curly) lines correspond to collinear quarks (gluons).}
\label{fig:Jq}
\eef
\bef
\includegraphics[width=4.2in]{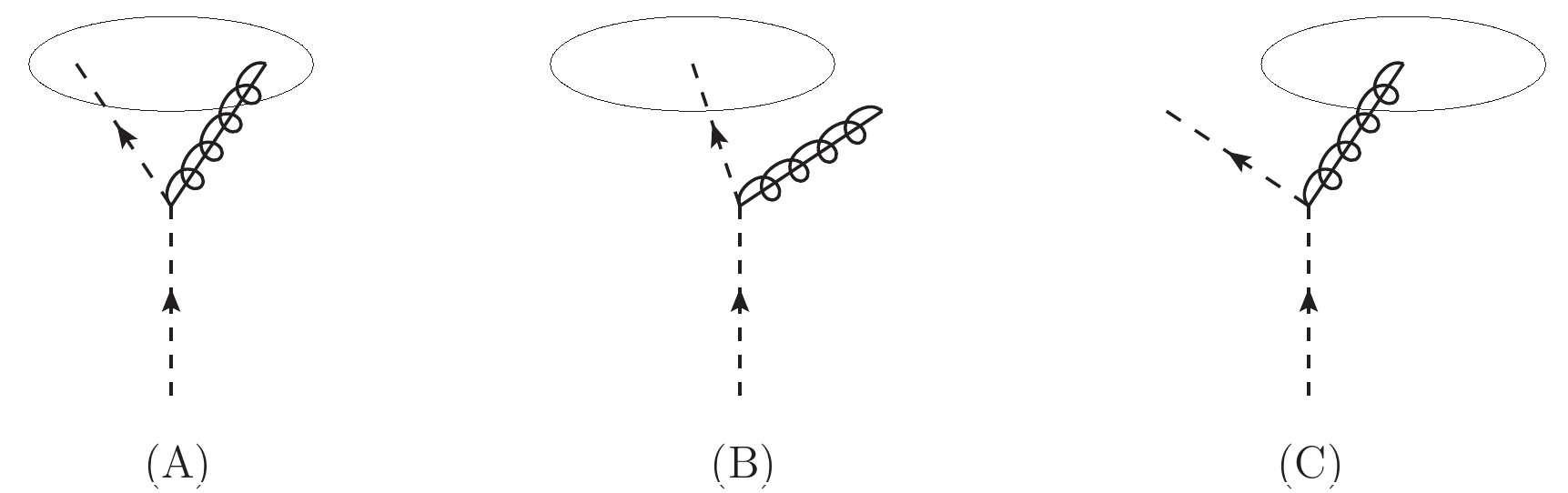}
\caption{The three contributions that need to be considered for the semi-inclusive quark fragmenting jet function: (A) both the quark and the gluon are inside the jet, (B) only the quark is inside the jet, (C) only the gluon is inside the jet.}
\label{fig:configuration}
\eef

At NLO, the SCET Feynman diagrams which contribute to the quark semi-inclusive FJF are shown in Fig.~\ref{fig:Jq}. See~\cite{Kang:2016mcy} for the corresponding diagrams for the gluon semi-inclusive FJF.
The semi-inclusive FJF is obtained by summing over all possible cuts of the Feynman diagrams shown in Fig.~\ref{fig:Jq}. We choose to work in pure dimensional regularization with $d=4-2\epsilon$ dimensions. We start by considering only cuts through loops, i.e. we only calculate the contributions where there are two final-state partons. The remaining cuts are virtual contributions leading to scaleless integrals which vanish in dimensional regularization. Effectively, virtual contributions only change IR poles to UV poles except for the IR poles that will eventually be matched onto the standard collinear fragmentation functions. In the end, we will be left with UV poles only, which will be subtracted by renormalization. 

As displayed in Fig.~\ref{fig:configuration}, for the quark semi-inclusive FJF there are two contributions that we need to consider (A) and (B)+(C), similar to the semi-inclusive jet function considered in~\cite{Kang:2016mcy}. To be specific, let us consider ${\cal G}_q^{q}(z, z_h,\omega_J)$. As displayed in Fig.~\ref{fig:Jq} -- the incoming quark has momentum $\ell^-$ and the final-state quark has momentum $\ell^- - q^-$. From these momenta  we can define the branching fraction $x=(\ell^- - q^-)/\ell^-$. At this order in perturbation theory there are only the following two possibilities. First, if both quark and gluon are inside the jet, as in Fig.~\ref{fig:configuration} (A), we have $z=1$ and $z_h=x$. Second, if the gluon exits the jet, as shown in Fig.~\ref{fig:configuration} (B), we have $z=x$ and $z_h=1$. Such considerations can be generalized to all other partonic splitting channels $i\to jk$, which we will now discuss in detail. 

\begin{enumerate}
\item
Both partons are inside the jet

This situation is shown in Fig.~\ref{fig:configuration}(A) for a quark initiated jet. In this case, all the initial quark energy $\omega$ is translated to the jet, hence, we have $z=\omega_J/\omega=1$. On the other hand, the energy of the fragmenting parton $\omega_h$ can be less than the jet energy and we will find in general for the ratio $z_h=\omega_h/\omega_J<1$. For a splitting process $i\to jk$, where $j$ denotes the fragmenting parton, the one-loop bare semi-inclusive FJF in the $\overline{\mathrm{MS}}$ scheme can be written as
\be\label{eq:GG1}
\GG_{i,\mathrm{bare}}^{jk,(1)}(z,z_h,\omega_J,\mu)=\delta(1-z)\f{\as}{\pi}\f{(e^{\gamma_E}\mu^2)^\epsilon}{\Gamma(1-\epsilon)}\hat P_{ji}(z_h,\epsilon)\int\f{dq_\perp}{q_\perp^{1+2\epsilon}}\Theta_{\mathrm{alg}} \, .
\ee
The superscript ``$jk$'' indicates that this is the ${\cal O}(\alpha_s)$ contribution where both partons $jk$ remain in the jet. The functions $\hat P_{ji}(z,\epsilon)$ are given by
\begin{subequations}
\label{eq:split1}
\bea
\hat P_{qq}(z, \epsilon) &= C_F\left[\frac{1+z^2}{1-z} - \epsilon\, (1-z) \right],
\\
\hat P_{gq}(z, \epsilon) &= C_F\left[\frac{1+(1-z)^2}{z} - \epsilon\, z \right],
\\
\hat P_{qg}(z, \epsilon) &= T_F\left[1-\frac{2z(1-z)}{1-\epsilon}\right],
\\
\hat P_{gg}(z, \epsilon) &= C_A\left[\frac{2z}{1-z} + \frac{2(1-z)}{z} + 2z(1-z) \right].
\eea
\end{subequations}
In the case when both partons are inside the jet, the jet algorithm constraints $\Theta_{\mathrm{alg}}$ for cone and \kt~algorithms with a jet radius $R$ are given in terms of the following theta functions:
\bea
\text{cone:~~} & \Theta_{\rm cone} = \theta\left((1-z_h)\omega_J \tan\frac{\R}{2} - q_\perp\right) \theta\left(z_h\omega_J \tan\frac{\R}{2} - q_\perp\right) \, ,
\label{eq:constraint1}
\\
\text{\kt:~~} & \Theta_{\text{\kt}} = \theta\left(z_h(1-z_h)\omega_J\tan\frac{\R}{2} - q_\perp\right) \, ,
\label{eq:constraint-kt}
\eea
where $\R$ is related to the jet radius $R$ as follows
\bea
\R \equiv \frac{R}{\cosh \eta},
\label{eq:R}
\eea
with $\eta$ the jet rapidity. For details, see \cite{Kang:2016mcy}.

Note that the expression in~(\ref{eq:GG1}) agrees with earlier work where the exclusive limit of the cross section was considered~\cite{Jain:2011xz,Chien:2015ctp}, except for the overall $\delta(1-z)$. In other words, part of the inclusive result is given by the exclusive FJF up to a trivial prefactor ensuring $z=1$. For the \kt~algorithm, the constraint in Eq.~\eqref{eq:constraint-kt} leads to the following $q_\perp$ integral
\be
\int  
\frac{dq_\perp}{q_\perp^{1+2\epsilon}} \Theta_{\text{\kt}} = \int_0^{z_h(1-z_h)\omega_J\tan\frac{\R}{2}}  \frac{dq_\perp}{q_\perp^{1+2\epsilon}} = - \frac{1}{2\epsilon} \left(\omega_J \tan\frac{\R}{2}\right)^{-2\epsilon} \left(z_h(1-z_h)\right)^{-2\epsilon}.
\ee
After substituting this expression into Eq.~(\ref{eq:GG1}), we obtain the contribution when both partons are inside the jet:
\bea
\label{eq:GGres1}
\GG_{i,\mathrm{bare}}^{jk,(1)}(z,z_h,\omega_J,\mu)=&\delta(1-z)\f{\as}{2\pi}\f{(e^{\gamma_E}\mu^2)^\epsilon}{\Gamma(1-\epsilon)}\hat P_{ji}(z_h,\epsilon)\left(-\f{1}{\epsilon}\right)
\nnu
&\times (z_h(1-z_h))^{-2\epsilon}\left(\omega_J\tan\f{\R}{2}\right)^{-2\epsilon} \, .
\eea

\item
Only one parton is inside the jet

The situation where one parton is inside and another parton is exiting the jet is illustrated in Fig.~\ref{fig:configuration}(B) and (C) for a quark initiated jet. In this case, the final-state quark (or gluon) forms the jet, with a jet energy $\omega_J = (\ell - q)^- = z\, \ell^-$. In other words, only a fraction $z$ of the incoming quark energy $\omega$ is translated into the jet energy. On the other hand, at this order in perturbation theory, all the jet energy is translated to the fragmenting parton inside the jet. Therefore, we will have an overall delta function ensuring $z_h=\omega_h/\omega_J=1$. It is important to contrast this situation with earlier work~\cite{Jain:2011xz,Chien:2015ctp}, where the exclusive limit of the FJF was considered. In that case, an upper cut $\Lambda$ for the total energy outside the measured jets was imposed to ensure the exclusive $n$-jet configuration. It was shown explicitly in the context of angularities in~\cite{Ellis:2010rwa} that for the exclusive case this contribution is power suppressed as ${\cal O}(\Lambda/Q)$, where $Q$ is the large scale of the process. In our case, when calculating an inclusive cross section, we do need to take into account this situation, as there is no constraint requiring that the energy of the exiting parton to be less than $\Lambda$. Instead, we need to integrate over all momentum configurations similar to the case of fragmentation functions. In turn, this implies that we do not have to impose anymore that the energy outside the reconstructed jets needs to be small, hence, we do not have power corrections of the form ${\cal O}(\Lambda/Q)$.

The constraints from the jet algorithms require that one of the partons is outside the jet, which can be written in the following form for both cone and \kt~algorithms
\bea
\Theta_{\rm cone} = \Theta_{\text{\kt}} = \theta\left(q_\perp - (1-z) \omega_J\tan\frac{\R}{2}\right) \, .
\label{eq:quarkin}
\eea
Note that this constraint is written in terms of $z$, whereas the constraints in Eq.~(\ref{eq:constraint1}) involve the variable $z_h$. We consider again the splitting process $i\to jk$, where only the parton $j$ remains inside the jet and eventually fragments into the observed hadron. We can express this part of the bare semi-inclusive FJF as
\be\label{eq:GG2}
\GG_{i,\mathrm{bare}}^{j(k),(1)}(z,z_h,\omega_J,\mu)=\delta(1-z_h)\f{\as}{\pi}\f{(e^{\gamma_E}\mu^2)^\epsilon}{\Gamma(1-\epsilon)}\hat P_{ji}(z,\epsilon)\int\f{dq_\perp}{q_\perp^{1+2\epsilon}}\Theta_{\mathrm{alg}} \, ,
\ee
where the superscript ``$j(k)$'' indicates that parton $k$ exits the jet. The structure here is very similar to Eq.~(\ref{eq:GG1}) except for the different overall delta function and a different jet algorithm constraint $\Theta_{\mathrm{alg}}$. We can now perform the $q_\perp$ integral as in~\cite{Kang:2016mcy}
\be
\int \frac{dq_\perp}{q_\perp^{1+2\epsilon}} \Theta_{\rm alg} = \int_{(1-z)\omega_J\tan\frac{R}{2}}^{\infty} \frac{dq_\perp}{q_\perp^{1+2\epsilon}} = \frac{1}{2\epsilon} \left(\omega_J \tan\frac{\R}{2}\right)^{-2\epsilon} \left(1-z\right)^{-2\epsilon} \, .
\ee
Eventually, we can write the second contribution of the bare semi-inclusive FJF as
\be\label{eq:GGres2}
\GG_{i,\mathrm{bare}}^{j(k),(1)}(z,z_h,\omega_J,\mu)=\delta(1-z_h)\f{\as}{2\pi}\f{(e^{\gamma_E}\mu^2)^\epsilon}{\Gamma(1-\epsilon)}\hat P_{ji}(z,\epsilon)\left(\f{1}{\epsilon}\right)(1-z)^{-2\epsilon}\left(\omega_J\tan\f{\R}{2}\right)^{-2\epsilon}\, .
\ee

\end{enumerate}

Adding the two contributions above in Eqs.~(\ref{eq:GGres1}) and~(\ref{eq:GGres2}), we obtain the following result for the ${\cal O}(\as)$ correction to the bare semi-inclusive FJF
\bea
\GG_{i,\mathrm{bare}}^{j}(z,z_h,\omega_J,\mu) = & \GG_{i,\mathrm{bare}}^{j,(0)}(z,z_h,\omega_J,\mu)  + \GG_{i,\mathrm{bare}}^{jk,(1)}(z,z_h,\omega_J,\mu) + \GG_{i,\mathrm{bare}}^{j(k),(1)}(z,z_h,\omega_J,\mu)
\nnu
= & \delta_{ij}\delta(1-z)\delta(1-z_h)+
\f{\as}{2\pi}\f{(e^{\gamma_E}\mu^2)^\epsilon}{\Gamma(1-\epsilon)}\left(-\f{1}{\epsilon}\right) \left(\omega_J\tan\f{\R}{2}\right)^{-2\epsilon}
\nnu
& \times \left[\delta(1-z)\hat P_{ji}(z_h,\epsilon) (z_h(1-z_h))^{-2\epsilon}-\delta(1-z_h)\hat P_{ji}(z,\epsilon)(1-z)^{-2\epsilon} \right]  \,.
\eea
Note that the leading-order result in Eq.~(\ref{eq:GLO}) only needs to be added for the case of $i=j$. As mentioned above, it can be seen here that the result to one-loop has always at least one delta function $\delta(1-z)$ or $\delta(1-z_h)$. However, this does not hold to higher orders in $\alpha_s$, where both $z,z_h$ can be smaller than one simultaneously.

We now continue by substituting  the explicit expressions for $\hat P_{ji}(z,\epsilon)$, as given in Eq.~(\ref{eq:split1}), and perform the expansion in powers of $\epsilon$. We obtain the following results for the semi-inclusive FJFs up to ${\cal O}(\alpha_s)$:
\bea
\label{eq:GGbareqq}
\GG_{q,\mathrm{bare}}^q(z,z_h,\omega_J,\mu)  = & \, \delta(1-z)\delta(1-z_h)+\f{\as}{2\pi}\left(-\f{1}{\epsilon}-L\right) P_{qq}(z_h)\delta(1-z) \nnu
& +\f{\as}{2\pi}\left(\f{1}{\epsilon}+L\right) P_{qq}(z)\delta(1-z_h) \nnu
& + \delta(1-z)\f{\as}{2\pi}\left[2C_F(1+z_h^2)\left(\f{\ln(1-z_h)}{1-z_h}\right)_+ + C_F(1-z_h)+2P_{qq}(z_h)\ln z_h \right] \nnu
& - \delta(1-z_h)\f{\as}{2\pi}\left[2C_F(1+z^2)\left(\f{\ln(1-z)}{1-z}\right)_+ +C_F(1-z)\right] \, ,
\\
\GG_{g,\mathrm{bare}}^g(z,z_h,\omega_J,\mu)  = & \, \delta(1-z)\delta(1-z_h)+\f{\as}{2\pi}\left(-\f{1}{\epsilon}-L\right) P_{gg}(z_h)\delta(1-z) \nnu
& +\f{\as}{2\pi}\left(\f{1}{\epsilon}+L\right) P_{gg}(z)\delta(1-z_h) \nnu
& + \delta(1-z)\f{\as}{2\pi}\left[4C_A\f{(1-z_h+z_h^2)^2}{z_h}\left(\f{\ln(1-z_h)}{1-z_h}\right)_+ +2P_{gg}(z_h)\ln z_h \right] \nnu
& - \delta(1-z_h)\f{\as}{2\pi}\left[4C_A\f{(1-z+z^2)^2}{z}\left(\f{\ln(1-z)}{1-z}\right)_+ \right] \, ,
\\
\GG_{q,\mathrm{bare}}^g(z,z_h,\omega_J,\mu)  = & \, \f{\as}{2\pi}\left(-\f{1}{\epsilon}-L\right) P_{gq}(z_h)\delta(1-z) +\f{\as}{2\pi}\left(\f{1}{\epsilon}+L\right) P_{gq}(z)\delta(1-z_h) \nnu
& + \delta(1-z)\f{\as}{2\pi}\left[2P_{gq}(z_h) \ln(z_h(1-z_h))+ C_F z_h \right] \nnu
& - \delta(1-z_h)\f{\as}{2\pi}\left[2P_{gq}(z)\ln(1-z)+C_F z \right] \, ,
\\
\label{eq:GGbaregq}
\GG_{g,\mathrm{bare}}^q(z,z_h,\omega_J,\mu)  = & \, \f{\as}{2\pi}\left(-\f{1}{\epsilon}-L\right) P_{qg}(z_h)\delta(1-z) +\f{\as}{2\pi}\left(\f{1}{\epsilon}+L\right) P_{qg}(z)\delta(1-z_h) \nnu
& + \delta(1-z)\f{\as}{2\pi}\left[2P_{qg}(z_h)\ln(z_h(1-z_h))+ 2T_F z_h(1-z_h) \right] \nnu
& - \delta(1-z_h)\f{\as}{2\pi}\left[2P_{qg}(z)\ln(1-z)+2T_F z(1-z) \right] \, .
\eea
Here, the logarithm $L$ is given by
\bea
\label{eq:L}
L=\ln\left(\f{\mu^2}{\omega_J^2\tan^2(\R/2)}\right),
\eea
and the functions $P_{ji}(z)$ are the usual Altarelli-Parisi splitting kernels
\begin{subequations}
\label{eq:split}
\bea
P_{qq}(z) &= C_F \left[ \frac{1+z^2}{(1-z)}_++\f{3}{2}\delta(1-z)\right]\, ,
\\
P_{gq}(z) &= C_F \frac{1+(1-z)^2}{z}\, ,
\\
P_{qg}(z) &= T_F(z^2 + (1-z)^2)\, ,
\\
P_{gg}(z) &= C_A\left[\frac{2z}{(1-z)}_+ + \frac{2(1-z)}{z} + 2z(1-z)\right]+\f{\beta_0}{2}\delta(1-z) \, ,
\eea
\end{subequations}
with $\beta_0=11/3 C_A-2/3 N_f$. The ``plus''-distributions are defined as usual via
\bea
\int_0^1dz \, f(z)[g(z)]_+\equiv \int_0^1 dz(f(z)-f(1))g(z)\,.
\eea
It is important to realize that the first poles (multiplied by $P_{ji}(z_h)\delta(1-z)$) for all the $\GG_{i,\mathrm{bare}}^ j$ in Eqs.~(\ref{eq:GGbareqq})-(\ref{eq:GGbaregq}) are IR poles that will be matched onto the standard collinear fragmentation functions as discussed below. On the other hand, the second poles (multiplied by $P_{ji}(z)\delta(1-z_h)$) in these results are UV poles after taking into account virtual corrections. Since the UV poles only involve the variable $z$, one should expect that the renormalization of the semi-inclusive FJF will also only involve the variable $z$. When dealing with the renormalization, the variable $z_h$ will only  be a parameter. However, since the IR poles involve only the variable $z_h$, it will be the relevant variable when matching onto the fragmentation functions. The renormalization and matching will be discussed below.

Note that we are only left with single poles $1/\epsilon$ and single logarithms $L$ here. All $1/\epsilon^2$ poles and $L^2$ terms that appear at intermediate steps of the calculations drop out. In fact, these terms cancel between the two contributions to the semi-inclusive FJF where both partons are in the jet and where one parton exits the jet. This is a crucial difference to the exclusive FJF~\cite{Jain:2011xz,Chien:2015ctp}. Basically, the exclusive FJF is given by the first part only, where both partons remain in the jet. In this case, there are $1/\epsilon^2$ poles as well as $L^2$ terms. The fact that here we only have single logarithms, whereas the exclusive case has double logarithms, leads to a very different renormalization and evolution of the exclusive and the semi-inclusive FJF as we are going to discuss in the next sections.

\subsection{Renormalization and RG evolution}

Our next step will be to renormalize the semi-inclusive FJF $\GG_i^j$ and, afterwards, to match onto the (also) renormalized partonic fragmentation functions in order to deal with the remaining IR divergences. The bare and renormalized semi-inclusive jet functions are related in the following way
\be
\GG_{i,\mathrm{bare}}^j(z,z_h,\omega_J,\mu)=\sum_k\int_z^1\f{dz'}{z'}\,Z_{ik}\left(\f{z}{z'},\mu\right)\,\GG_k^j(z',z_h,\omega_J,\mu) \, ,
\ee
where $Z_{ik}(z/z',\mu)$ is the renormalization matrix. We would like to emphasize that only the variable $z$ is involved in this convolution and, hence, in the renormalization, since the UV poles only involve the variable $z$ as demonstrated in last section. On the other hand, the variable $z_h$ is only a parameter here, but it will be the relevant variable when matching onto the fragmentation functions. The renormalized FJF satisfies the following RG evolution equation
\be
\mu\f{d}{d\mu}\GG_i^j(z,z_h,\omega_J,\mu)=\sum_k\int_z^1\f{dz'}{z'}\gamma_{ik}^\GG\left(\f{z}{z'},\mu\right)\,\GG_k^j(z',z_h,\omega_J,\mu) \, ,
\ee
where the anomalous dimension matrix is given by
\be
\gamma_{ij}^\GG=-\sum_k\int_z^1\f{dz'}{z'}\left(Z\right)^{-1}_{ik}\left(\f{z}{z'},\mu\right)\,\mu\f{d}{d\mu}Z_{kj}\left(z',\mu\right) \, .
\ee
The inverse of the renormalization matrix is defined via
\bea
\sum_k \int_z^1 \frac{dz'}{z'} \left(Z\right)^{-1}_{ik}\left(\frac{z}{z'}, \mu\right)  Z_{kj}(z', \mu)=\delta_{ij} \delta(1-z).
\eea
For the renormalization matrix up to ${\cal O}(\as)$, we find
\bea
Z_{ij}(z,\mu) = \delta_{ij}\delta(1-z) + \frac{\alpha_s(\mu)}{2\pi} \left(\frac{1}{\epsilon}\right) P_{ji}(z) \, ,
\eea
and, hence, the anomalous dimension matrix is given by
\bea
\gamma_{ij}^\GG(z, \mu) =  \frac{\alpha_s(\mu)}{\pi}  P_{ji}(z) \, .
\eea
This implies that the renormalized semi-inclusive FJF follows the usual timelike DGLAP evolution equation for fragmentation functions~\cite{Gribov:1972ri, Lipatov:1974qm, Dokshitzer:1977sg,Altarelli:1977zs}
\be\label{eq:DGLAP1}
\mu \frac{d}{d\mu} \GG_i^h(z,z_h, \omega_J, \mu) = \frac{\alpha_s(\mu)}{\pi} \sum_k \int_z^1  \frac{dz'}{z'} P_{ki}\left(\frac{z}{z'} \right) \GG_k^h(z',z_h, \omega_J,\mu) \, ,
\ee
where we have switched back from the semi-inclusive partonic FJF to the hadronic FJF. An analogous result was found for the semi-inclusive jet function in~\cite{Kang:2016mcy}. The leading-order evolution kernels $P_{ji}(z)$ were defined in~(\ref{eq:split}). We would like to again contrast this finding  to the exclusive limit of the FJF as considered in~\cite{Jain:2011xz,Chien:2015ctp}. The exclusive FJF satisfies a multiplicative RG equation and its anomalous dimension has a logarithmic dependence. Solving this RG equation leads to the exponentiation of double logarithms $\as^n\ln^{2n} R$. In our case, we have the DGLAP convolution structure and its solution, as discussed below, will lead to the resummation of single logarithms $\as^n\ln^n R$.

For completeness, we list all four renormalized semi-inclusive partonic FJFs here for the \kt~algorithm
\begin{subequations}
\label{eq:GGrenqq}
\bea
\GG_{q}^q(z,z_h,\omega_J,\mu)  = & \, \delta(1-z)\delta(1-z_h)+\f{\as}{2\pi}\left(-\f{1}{\epsilon}-L\right) P_{qq}(z_h)\delta(1-z) +\f{\as}{2\pi} L\, P_{qq}(z)\delta(1-z_h) 
\nnu
& + \delta(1-z)\f{\as}{2\pi}\left[2C_F(1+z_h^2)\left(\f{\ln(1-z_h)}{1-z_h}\right)_+ + C_F(1-z_h)+2P_{qq}(z_h)\ln z_h \right] \nnu
& - \delta(1-z_h)\f{\as}{2\pi}\left[2C_F(1+z^2)\left(\f{\ln(1-z)}{1-z}\right)_+ +C_F(1-z) \right] \, ,
\\
\GG_{g}^g(z,z_h,\omega_J,\mu)  = & \, \delta(1-z)\delta(1-z_h)+\f{\as}{2\pi}\left(-\f{1}{\epsilon}-L\right) P_{gg}(z_h)\delta(1-z) +\f{\as}{2\pi} L\, P_{gg}(z)\delta(1-z_h) \nnu
& + \delta(1-z)\f{\as}{2\pi}\left[4C_A\f{(1-z_h+z_h^2)^2}{z_h}\left(\f{\ln(1-z_h)}{1-z_h}\right)_+ +2P_{gg}(z_h)\ln z_h \right] \nnu
& - \delta(1-z_h)\f{\as}{2\pi}\left[4C_A\f{(1-z+z^2)^2}{z}\left(\f{\ln(1-z)}{1-z}\right)_+ \right] \, ,
\\
\GG_{q}^g(z,z_h,\omega_J,\mu)  = & \, \f{\as}{2\pi}\left(-\f{1}{\epsilon}-L\right) P_{gq}(z_h)\delta(1-z) +\f{\as}{2\pi}L\, P_{gq}(z)\delta(1-z_h) \nnu
& + \delta(1-z)\f{\as}{2\pi}\left[2P_{gq}(z_h)\ln(z_h(1-z_h))+ C_F z_h \right] \nnu
& - \delta(1-z_h)\f{\as}{2\pi}\left[2P_{gq}(z)\ln(1-z)+z \right] \, ,
\\
\GG_{g}^q(z,z_h,\omega_J,\mu)  = & \, \f{\as}{2\pi}\left(-\f{1}{\epsilon}-L\right) P_{qg}(z_h)\delta(1-z) +\f{\as}{2\pi}L\, P_{qg}(z)\delta(1-z_h) \nnu
& + \delta(1-z)\f{\as}{2\pi}\left[2P_{qg}(z_h) \ln(z_h(1-z_h))+ 2T_F z_h(1-z_h) \right] \nnu
& - \delta(1-z_h)\f{\as}{2\pi}\left[2P_{qg}(z)\ln(1-z)+2T_F z(1-z) \right] \, .
\eea
\end{subequations}
The remaining poles here are IR poles which we are going to match onto the collinear fragmentation functions in the next section.

\subsection{Matching onto standard collinear fragmentation functions}

At a  scale $\mu\gg\Lambda_{\mathrm{QCD}}$, we can match the semi-inclusive FJF $\GG_i^j(z,z_h,\omega_J,\mu)$ onto the fragmentation functions $D_i^h(z,\mu)$ as follows:
\bea
\label{eq:matching}
\GG_i^h(z,z_h,\omega_J,\mu) = \sum_j \int_{z_h}^1 \frac{dz_h'}{z_h'} {\mathcal J}_{ij}\left(z,z_h',\omega_J,\mu\right) D_j^h\left(\frac{z_h}{z_h'},\mu\right) \, ,
\eea
a relation valid up to the power correction of ${\mathcal O}(\Lambda^2_{\rm QCD}/\omega^2\tan^2(\R/2))$~\cite{Jain:2011xz,Chien:2015ctp}. Note that the convolution variable here is $z_h$, whereas $z$ is a mere parameter. Other than that, this procedure is completely analogous to the case of exclusive FJF~\cite{Jain:2011xz,Chien:2015ctp}. To obtain the matching coefficients ${\mathcal J}_{ij}$, we replace the hadron $h$ by a parton state, using the perturbative results for the renormalized $\GG_i^j(z,z_h,\omega_J,\mu)$ and $D_i^j(z_h,\mu)$. While the $\GG_i^j(z,z_h,\omega_J,\mu)$ are given in Eq.~\eqref{eq:GGrenqq}, the perturbative renormalized standard collinear fragmentation functions $D_i^j(z_h,\mu)$ up to ${\cal O}(\alpha_s)$ using pure dimensional regularization in the $\overline{\mathrm{MS}}$ scheme are given by
\be
D_{i}^j(z_h, \mu) = \delta_{ij}\delta(1-z_h) + \frac{\alpha_s}{2\pi}P_{ji}(z_h)  \left(-\frac{1}{\epsilon}\right)\,.
\ee
Finally, the matching coefficients ${\cal J}_{ij}(z,z_h,\omega_J,\mu)$ for both \kt~and cone algorithms are given by
\bea
\label{eq:nloresult}
{\cal J}_{qq}(z,z_h,\omega_J,\mu) = &\delta(1-z)\delta(1-z_h)+\f{\as}{2\pi}\Bigg\{L\,\left[P_{qq}(z)\delta(1-z_h) - P_{qq}(z_h)\delta(1-z) \right] \nnu
& +\, \delta(1-z)\left[2C_F(1+z_h^2)\left(\f{\ln(1-z_h)}{1-z_h}\right)_+ + C_F(1-z_h)+{\cal I}^{\mathrm{alg}}_{qq}(z_h) \right]
\nnu
& -\,  \delta(1-z_h) \left[2C_F(1+z^2)\left(\f{\ln(1-z)}{1-z}\right)_+ + C_F(1-z) \right]     \Bigg\} \, ,
\nnu
{\cal J}_{gg}(z,z_h,\omega_J,\mu) = &\delta(1-z)\delta(1-z_h)+\f{\as}{2\pi}\Bigg\{L\,\left[P_{gg}(z)\delta(1-z_h) - P_{gg}(z_h)\delta(1-z) \right]\nnu
& +\, \delta(1-z)\left[4C_A\f{(1-z_h+z_h^2)^2}{z_h}\left(\f{\ln(1-z_h)}{1-z_h}\right)_+ +{\cal I}^{\mathrm{alg}}_{gg}(z_h) \right]
\nnu
& -\,  \delta(1-z_h) \left[4C_A\f{(1-z+z^2)^2}{z}\left(\f{\ln(1-z)}{1-z}\right)_+ \right]     \Bigg\} \, ,
\nnu
{\cal J}_{qg}(z,z_h,\omega_J,\mu)  = & \f{\as}{2\pi}\Bigg\{ L\,\left[P_{gq}(z)\delta(1-z_h) - P_{gq}(z_h)\delta(1-z) \right]\nnu
& +\, \delta(1-z)\left[2P_{gq}(z_h)\ln(1-z_h)+C_F z_h+{\cal I}_{qg}^\mathrm{alg}(z_h) \right] \nnu
& -\, \delta(1-z_h)\left[2P_{gq}(z)\ln(1-z)+C_F z \right] \Bigg\}\, , 
\nnu
{\cal J}_{gq}(z,z_h,\omega_J,\mu)  = &\f{\as}{2\pi}\Bigg\{L\,\left[P_{qg}(z)\delta(1-z_h) - P_{qg}(z_h)\delta(1-z) \right]\nnu
& +\, \delta(1-z)\left[2P_{qg}(z_h)\ln(1-z_h)+2T_F z_h(1-z_h)+{\cal I}_{gq}^\mathrm{alg}(z_h) \right] \nnu
& -\, \delta(1-z_h)\left[2P_{qg}(z)\ln(1-z)+2T_F z(1-z) \right] \Bigg\} \, ,
\eea
where ${\cal I}^{\mathrm{alg}}_{ij}(z_h)$ are jet algorithm-dependent functions with the following expressions
\bea
{\cal I}_{ij}^{\mathrm{anti-k}_T}(z_h)  = & 2 P_{ji}(z_h)\,\ln z_h \, , \\
{\cal I}_{ij}^{\mathrm{cone}}(z_h) = & 2P_{ji}(z_h)\,\ln(z_h/(1-z_h))\theta(1/2-z_h) \, .
\eea
Note that at this order in perturbation theory only the terms ${\cal J}_{ij}(z,z_h,\omega_J,\mu)\sim\delta(1-z)$ have jet algorithm-dependent contributions. This can be understood in the sense that up to ${\cal O}(\as)$, all terms $\sim\delta(1-z_h)$ correspond to a configuration where there is only one parton inside the jet and the jet algorithm constraints for both cone and \kt~jets are the same, cf. Eq.~(\ref{eq:quarkin}). After taking into account the different conventions used in this work and the formalism developed in~\cite{Kaufmann:2015hma}, where the NLO calculation was performed, we find full agreement between the two approaches at this level. However, having obtained those results within the effective theory framework, we can go beyond the fixed order calculation. With the help of the derived DGLAP renormalization group equations we can now resum single-logarithms in $R$, which will be discussed in the next section. Before we do that, it is useful to remind the reader that we do not assume any hierarchy between $z$ and $z_h$. Recall that $z$ is the energy of the jet compared to the initiating parton energy, whereas $z_h$ is the energy of the hadron divided by the jet energy. As can be seen form the above equations, at leading-order, we have both $z=z_h=1$. At NLO, either $z=1$ or $z_h=1$. Beyond NLO both $z$ and $z_h$ can be smaller than one simultaneously, but there is no hierarchy between them.

The first step toward resummation, discussed in the following section, is to choose a scale $\mu$ for the matching coefficients listed above, such that it minimizes the large logarithms. From the explicit expressions, one finds that an obvious choice is
\bea
\mu=\mu_\GG=\omega_J\tan(\R/2) \, ,
\eea
which sets the logarithms $L$ as defined in~(\ref{eq:L}) to zero. It might be instructive to realize that
\bea
\mu_\GG=\omega_J\tan(\R/2) =  \left(2 p_T \cosh \eta \right)\tan\left(\frac{R}{2\cosh \eta}\right) \approx p_T R \equiv p_{TR},
\eea
where we have used Eq.~\eqref{eq:R} and the fact that the jet energy is $\omega_J = 2 p_T \cosh \eta$. Evolving the semi-inclusive FJF from this scale $\mu_\GG\sim p_{TR}$ to the hard scale of the process $\mu\sim p_T$, we are thus resumming the logarithms in $R$. Furthermore, we would like to point out that the derived matching coefficients satisfy certain sum rules after integrating over the fragmenting parton variable $z_h$, as also pointed out in~\cite{Jain:2011xz,Kaufmann:2015hma}. By summing over the quark and gluon initiated contributions respectively and integrating over $z_h$, we obtain the semi-inclusive jet functions $J_{q,g}(z,\omega_J,\mu)$ introduced in~\cite{Kang:2016mcy}
\bea
\int_0^1dz_h\,z_h\big[{\cal J}_{qq}(z,z_h,\omega_J,\mu)+{\cal J}_{qg}(z,z_h,\omega_J,\mu)\big] &= J_q(z,\omega_J,\mu) \, ,
\\
\int_0^1dz_h\,z_h\big[{\cal J}_{gg}(z,z_h,\omega_J,\mu)+2N_f{\cal J}_{gq}(z,z_h,\omega_J,\mu)\big] &= J_g(z,\omega_J,\mu) \, .
\eea

\subsection{$\ln R$ Resummation}

We define the following renormalized and matched semi-inclusive FJFs for quarks and gluons
\begin{subequations}
\label{eq:GGqg}
\bea
\GG_q^h(z,z_h,\omega_J,\mu_\GG)  = & \int_{z_h}^1\f{dz_h'}{z_h'}\left[{\cal J}_{qq}(z,z_h',\omega_J,\mu_\GG) D_q^h\left(\f{z_h}{z_h'},\mu_\GG\right) \right. \nnu
&\left. +{\cal J}_{qg}(z,z_h',\omega_J,\mu_\GG)D_g^h\left(\f{z_h}{z_h'},\mu_\GG\right)\right] \, , 
\\
\GG_g^h(z,z_h,\omega_J,\mu_\GG)  = &\int_{z_h}^1\f{dz_h'}{z_h'}\left[{\cal J}_{gg}(z,z_h',\omega_J,\mu_\GG) D_g^h\left(\f{z_h}{z_h'},\mu_\GG\right) \right. \nnu
&\left. +\sum_{i=q,\bar q} {\cal J}_{gi}(z,z_h',\omega_J,\mu_\GG)D_i^h\left(\f{z_h}{z_h'},\mu_\GG\right)\right] \, ,  
\eea
\end{subequations}
which constitute the initial conditions for the DGLAP equations at scale $\mu_\GG\sim p_{TR}$. Following Eq.~(\ref{eq:DGLAP1}), the timelike DGLAP evolution equations for the semi-inclusive FJF can be cast into the following form
\be\label{eq:DGLAP2}
\frac{d}{d \log \mu^2}
\begin{pmatrix}
\GG^h_S (z,z_h,\omega_J,\mu) \\ \GG^h_g(z,z_h,\omega_J,\mu)
\end{pmatrix}
=
\frac{\alpha_s(\mu)}{2 \pi}
\begin{pmatrix}
P_{qq}(z) & ~2 N_f P_{gq}(z) \\
P_{qg}(z) & ~P_{gg}(z)
\end{pmatrix}
\otimes
\begin{pmatrix}
\GG^h_S (z,z_h,\omega_J,\mu)\\ \GG^h_g(z,z_h,\omega_J,\mu)
\end{pmatrix},
\ee
where the $P_{ji}(z)$ denote the leading-order Altarelli-Parisi splitting kernels as given in~(\ref{eq:split}) and $\otimes$ denotes the usual convolution which is here taken in the variable $z$ only
\bea
(f\otimes g)(z)=\int_z^1\f{dz'}{z'} f(z')g(z/z') \, .
\eea
It is important to emphasize again that the above DGLAP equations only evolve the semi-inclusive FJFs in $(z, \mu)$ space, while the $z_h$-dependence in $\GG_{q,g}^h(z,z_h,\omega_J,\mu)$ is completely determined by the initial conditions in Eq.~\eqref{eq:GGqg} and remains unchanged in the evolution. The function $\GG_S^h(z,z_h,\omega_J,\mu)$ in~(\ref{eq:DGLAP2}) is the singlet semi-inclusive FJF given by the sum over all quarks and anti-quarks
\be
\GG_S^h(z,z_h,\omega_J,\mu)=\sum_{i=q,\bar q}\GG_i^h(z,z_h,\omega_J,\mu) \, .
\ee
Note that here we also need to consider a separate non-singlet evolution as the $\GG_i^h(z,z_h,\omega_J,\mu)$ depend on fragmentations functions that can be different for all quark flavors $i=q,\bar q$, i.e. due to the difference in $z_h$-dependence. This is different to the semi-inclusive jet function $J_i(z, \omega_J,\mu)$ considered in~\cite{Kang:2016mcy} which is purely perturbative and depends on $z$ only, and thus it is the same for all quark flavors. We follow the conventions of~\cite{Vogt:2004ns} for performing the separate non-singlet evolution.

The initial condition for the DGLAP equations at the scale $\mu_\GG$ involve delta functions and ``plus'' distributions. Therefore, we solve the evolution equations in Mellin moment space following the method of~\cite{Vogt:2004ns}. The Mellin moments of any $z$-dependent function are defined as
\be
f(N)=\int_0^1 dz\, z^{N-1} f(z) \, .
\ee
The Mellin moments of delta functions and ``plus'' distributions are simple functions in Mellin moment space. This way, we can perform the evolution from the initial scale $\mu_\GG$ to any scale $\mu$ in Mellin space and take an inverse transformation afterwards, which is given by a contour integral in the complex $N$ plane as defined below~(\ref{eq:MellinInverse}). An advantage of the treatment in Mellin moment space is that the convolution structure of the DGLAP equations in~(\ref{eq:DGLAP2}) is turned into simple products in Mellin space. In general, one has schematically
\be
(f\otimes g)(N)=f(N)\, g(N) \, .
\ee
Taking into account the fact that the singlet evolution matrix in~(\ref{eq:DGLAP2}) only depends on the scale $\mu$ through the strong coupling constant $\as(\mu)$, we can write down the leading-order solution for the timelike DGLAP evolution equation for the semi-inclusive FJFs as~\cite{Vogt:2004ns}
\ba\label{eq:DGLAP3}
\begin{pmatrix}
\GG^h_S (N,z_h,\omega_J,\mu) 
\\ \GG^h_g (N,z_h,\omega_J,\mu)
\end{pmatrix}
& = &
\left[
e_+(N)
\left(\frac{\alpha_s(\mu)}{\alpha_s(\mu_\GG)} \right)^{-r_-(N)}
+ e_-(N)
\left(\frac{\alpha_s(\mu)}{\alpha_s(\mu_\GG)} \right)^{-r_+(N)}
\right] 
\nnu
&& \times
\begin{pmatrix}
 \GG^h_S (N,z_h,\omega_J,\mu_\GG) \\  \GG^h_g (N,z_h,\omega_J,\mu_\GG)
\end{pmatrix} 
\, ,
\ea
where $r_+(N)$ and $r_-(N)$ denote the larger and smaller eigenvalue of the leading-order singlet evolution matrix, see~(\ref{eq:DGLAP2}),
\be
r_{\pm}(N)=\f{1}{2\beta_0}\left[P_{qq}(N)+P_{gg}(N)\pm\sqrt{\left(P_{qq}(N)-P_{gg}(N)\right)^2 +4 P_{qg}(N) P_{gq}(N)} \right] \, .
\ee
The projector matrices $e_{\pm}(N)$ in~(\ref{eq:DGLAP3}) are defined as
\be
e_{\pm}(N)=\f{1}{r_\pm(N) - r_\mp(N)}
\begin{pmatrix}
P_{qq}(N)-r_\mp(N) & ~2 N_f P_{gq}(N) \\
P_{qg}(N) & ~P_{gg}(N)-r_\mp(N)
\end{pmatrix}
 \, .
\ee
The evolved semi-inclusive FJFs in $z$-space can be obtained by performing a Mellin inverse transformation
\be\label{eq:MellinInverse}
\GG^h_{S,g}(z,z_h,\omega_J,\mu) = \f{1}{2\pi i} \int_{{\cal C}_N} dN\, z^{-N} \GG^h_{S,g}(N,z_h,\omega_J,\mu)\, ,
\ee
where the contour in the complex $N$ plane is chosen to the right of all the poles in $\GG^h_{S,g}(N,z_h,\omega_J,\mu)$. Note that the outlined solution here only leads to a leading-logarithmic (LL$_R$) resummation since we evolve the semi-inclusive FJF using the leading-order splitting kernels. However, it is straightforward to extend the solution of the evolution equations to NLL$_R$ accuracy. Using NLO splitting kernels and by expanding the solution in Mellin space around its leading-order solution, we may directly achieve a combined precision of NLO+NLL$_R$. See~\cite{Vogt:2004ns,Anderle:2015lqa,Kang:2016mcy} for more detailed discussions. In~\cite{Kang:2016mcy}, it was shown that the difference between LL$_R$ and NLL$_R$ is only of the order of a few percent for the inclusive jet production cross section. The difference is even less significant for the JFF which is given by a ratio of two (LL$_R$ or NLL$_R$) resummed quantities, see Eq.~(\ref{eq:JFFdef}) below. Nevertheless, for consistency with the non-perturbative fragmentation functions that are evolved using NLO splitting kernels, we choose to evolve the semi-inclusive FJFs also at NLL$_R$ accuracy. In addition, similar to hadron-inclusive cross sections, it is consistent to use the NLO splitting kernels in order to fully capture all ${\cal O}(\alpha_s^3)$ effects.

Our evolution code~\cite{Kang:2016mcy} is a modified version of the evolution code for fragmentation functions presented in~\cite{Anderle:2015lqa}, which in turn is based on the {\sc Pegasus} evolution package for PDFs~\cite{Vogt:2004ns}. We would like to emphasize again that we have to solve two DGLAP equations now. The fragmentation functions $D_i^h(z_h, \mu)$ describing the parton-to-hadron fragmentation inside the jet enters the semi-inclusive FJF through Eq.~\eqref{eq:GGqg}, and are evolved from their initial scale, $\sim 1$~GeV, up to the scale $\mu_\GG\sim p_{TR}$ using the standard timelike DGLAP equations. This part of evolution is directly associated with the variable $z_h$. In addition, the semi-inclusive FJFs are evolved as a whole from $\mu_\GG\sim p_{TR}$ to $\mu\sim p_T$ using also the same timelike DGLAP evolution equations, though the evolution now is directly tied to the variable $z$ in Eq.~\eqref{eq:DGLAP2}. 

\section{Phenomenology for $pp\to ({\rm jet}\, h)X$ \label{sec:three}}

Following~\cite{Kaufmann:2015hma,Kang:2016mcy}, we can write the cross section for $pp\to (\mathrm{jet}h)X$ as
\bea
\label{eq:sigjethX}
\frac{d\sigma^{pp\to (\mathrm{jet}h)X}}{dp_Td\eta dz_h}  = & \frac{2 p_T}{s}\sum_{a,b,c}\int_{x_a^{\mathrm{min}}}^1\f{dx_a}{x_a}f_a(x_a,\mu)\int_{x_b^{\mathrm{min}}}^1\f{dx_b}{x_b} f_b(x_b,\mu)  
\nnu
&\times 
\int^1_{z_c^{\mathrm{min}}} \frac{dz_c}{z_c^2}\frac{d\hat\sigma^c_{ab}(\hat s,\hat p_T,\hat \eta,\mu)}{dvdz}\GG_c^h(z_c,z_h,\omega_J,\mu) \; ,
\eea
where $\sum_{a,b,c}$ stands for a sum over all parton flavors, and $f_{a/b}(x_{a/b},\mu)$ are the usual parton distribution functions. On the other hand, the relevant semi-inclusive FJFs $\GG_{q,g}^h(z,z_h,\omega_J,\mu)$ are evolved from their natural scale $\mu_\GG\sim p_{TR}$ to the scale $\mu$ as in~(\ref{eq:DGLAP3}). Here, $s$, $p_T$ and $\eta$ correspond to the center-of-mass  (CM) system energy, the jet transverse momentum and the jet rapidity, respectively. The hard functions $d\hat\sigma_{ab}^c(\hat s,\hat p_T,\hat\eta,\mu)$ are functions of the corresponding partonic variables: the partonic CM energy $\hat s=x_ax_bs$, the partonic transverse momentum $\hat p_T=p_T/z_c$ and the partonic rapidity $\hat\eta=\eta-\ln(x_a/x_b)/2$. The variables $v,z$ can be expressed in terms of these partonic variables
\be
v=1-\f{2\hat p_T}{\sqrt{\hat s}}e^{-\hat\eta}, \qquad z=\f{2\hat p_T}{\sqrt{s}}\cosh\hat\eta \, .
\ee
Up to ${\cal O}(\as)$, the hard functions take the form
\bea
\frac{d\hat\sigma_{ab}^c}{dvdz} = \frac{d\hat\sigma_{ab}^{c,(0)}}{dv}\delta(1-z)+\frac{\alpha_s(\mu)}{2\pi} \frac{d\hat\sigma_{ab}^{c,(1)}}{dvdz}.
\eea
As pointed out in~\cite{Kang:2016mcy}, the hard functions here are the same as the partonic cross sections for the processes $pp\to\mathrm{jet}X$ and $pp\to hX$, see also~\cite{Fickinger:2016rfd}. The corresponding expressions were presented in~\cite{Aversa:1988vb,Jager:2002xm}. Finally, the integration limits in~(\ref{eq:sigjethX}) are customarily written in terms of the hadronic variables $V,Z$,
\be
V=1-\f{2p_T}{\sqrt{s}}e^{-\eta}, \qquad Z=\f{2 p_T}{s}\cosh\eta \, ,
\ee
and are given by
\be
x_a^{\mathrm{min}}=1-\f{1-Z}{V},\quad x_b^\mathrm{min}=\f{1-V}{1+(1-V-Z)/x_a},\quad z_c^\mathrm{min}=\f{1-V}{x_b}-\f{1-V-Z}{x_a} \, .
\ee
The corresponding inclusive jet production cross section $pp\to\mathrm{jet}X$ including $\ln R$ resummation was derived in~\cite{Kang:2016mcy} and is given by
\bea\label{eq:sigjetX}
\frac{d\sigma^{pp\to \mathrm{jet}X}}{dp_Td\eta} =& \frac{2 p_T}{s}\sum_{a,b,c}\int_{x_a^{\mathrm{min}}}^1\f{dx_a}{x_a}f_a(x_a,\mu)\int_{x_b^{\mathrm{min}}}^1\f{dx_b}{x_b} f_b(x_b,\mu) 
\nnu
&\times
\int^1_{z_c^{\mathrm{min}}} \frac{dz_c}{z_c^2} \frac{d\hat\sigma^c_{ab}(\hat s,\hat p_T,\hat \eta,\mu)}{dvdz}J_c(z_c,\omega_J,\mu) \, ,
\eea
where $J_c(z_c,\omega_J,\mu)$ is the semi-inclusive jet function introduced in~\cite{Kang:2016mcy}. It was shown in~\cite{Kang:2016mcy} that $J_c(z_c,\omega_J,\mu)$ follows a similar DGLAP type evolution from the initial scale $\mu_J\sim p_{TR}$ to the hard scale $\mu$, which resums single-logarithms $\ln R$. Here $\mu_J$ is the characteristic scale for semi-inclusive jet function $J_c(z_c,\omega_J,\mu)$, as shown in~\cite{Kang:2016mcy}. As mentioned above, the hard functions here, $d\hat\sigma_{ab}^c$, are the same as in Eq.~(\ref{eq:sigjethX}). In addition, all kinematic variables and integration limits are the same as in Eq.~(\ref{eq:sigjethX}). Note that power corrections to both factorization formulas in Eqs.~(\ref{eq:sigjethX}) and~(\ref{eq:sigjetX}) are of the order of ${\cal O}(\Lambda_{\mathrm{QCD}}^2/\omega^2)$ in the massless case we consider, see e.g.~\cite{Collins:1989gx}. In addition, there are power corrections that are purely of kinematic origin, such as hadron mass corrections,  which lead to corrections of ${\cal O}(m_h^2/\omega^2)$, where $m_h$ denotes the hadron mass. The JFF, as measured by experiments is given by the following ratio
\bea
\label{eq:JFFdef}
F(z_h,p_T)=\frac{d\sigma^{pp\to (\mathrm{jet}h)X}}{dp_Td\eta dz_h}\Big/ \frac{d\sigma^{pp\to \mathrm{jet}X}}{dp_Td\eta},
\eea
where $z_h=p_T^h/p_T$, $p_T$ and $\eta$ are integrated over certain bins. We choose to match the $\ln R$ resummation onto the fixed NLO calculation in the sense that we do not take into account ${\cal O}(\as^4)$ contributions that appear in the convolution of the hard functions $d\hat\sigma_{ab}^c$ and the semi-inclusive FJF $\GG_c^h$. Schematically, we have
\bea
\label{eq:NLOmatching}
\left(d\hat\sigma^{c,(0)}_{ab} + d\hat\sigma^{c,(1)}_{ab} \right)\otimes\left( \GG_c^{h,(0)}+\GG_c^{h,(1)} \right) 
= &\left(d\hat\sigma^{c,(0)}_{ab} + d\hat\sigma^{c,(1)}_{ab} \right)\otimes\, \GG_c^{h,(0)} + d\hat\sigma^{c,(0)}_{ab}\otimes\, \GG_c^{h,(1)} 
\nnu
&+ {\cal O}(\as^4)\, ,
\eea
where the term $d\hat\sigma^{c,(1)}_{ab}\otimes\, \GG_c^{h,(1)}$ is at the order of ${\cal O}(\as^4)$, i.e., part of NNLO contributions, and will be dropped for consistency. This way, we get back to the fixed NLO calculation of~\cite{Kaufmann:2015hma} in the limit of no evolution $\mu_\GG\to\mu$ for the semi-inclusive FJF. Since the initial scale of the evolution depends on $R$, we obtain the limit of no evolution for $R\to 1$. Even though the limit of no evolution, $R\to 1$, is beyond the approximation of narrow jets, it serves as an important numerical check of our DGLAP based resummation code. As can be seen from~Eq.~\eqref{eq:NLOmatching}, we need to evolve all $\GG_i^{h,(0)}$ and $\GG_i^{h,(1)}$ separately in order to achieve the correct matching at NLO. Numerically, this procedure is particularly challenging as the initial condition at leading-order is $\GG_c^{h,(0)}\sim\delta(1-z)$. Also the one-loop semi-inclusive FJF $\GG_c^{h,(1)}$ involve delta functions and ``plus'' distributions at the initial scale $\mu_\GG$. Even with a long evolution $\mu_\GG\to\mu$, the evolved functions are divergent for $z\to 1$ and can not be simply substituted into the expression for the cross section in~Eq.~\eqref{eq:sigjethX}. We deal with this issue by adopting a procedure developed in~\cite{Bodwin:2015iua} in the context of the evolution of fragmentation functions for quarkonia. For more details, we refer the reader to~\cite{Kang:2016mcy}, where the same procedure was adopted for the semi-inclusive jet function in the application for $pp\to\mathrm{jet}X$.

\subsection{Comparison to LHC data}

We will now  compare the numerical results obtained within the new formalism at NLO+NLL$_R$ to experimental data on light charged hadrons inside a jet in proton-proton collisions at the LHC. See also~\cite{Kaufmann:2015hma,Chien:2015ctp} for detailed discussions. For all data sets presented here, the \kt~algorithm was used to reconstruct the jets. The ATLAS experiment has published data for the JFF at a CM energy of $\sqrt{s}=7$~TeV~\cite{Aad:2011sc} and $\sqrt{s}=2.76$~TeV~\cite{ATLAS:2015mla}. For the data at $\sqrt{s}=7$~TeV, the ATLAS experiment uses a jet radius parameter  $R=0.6$, the jet rapidity is integrated over the interval  $|\eta|<1.2$, and the jet transverse momentum $p_T$ is divided into several bins ranging from 25 to $500$~GeV. On the other hand, for $\sqrt{s}=2.76$~TeV, a jet radius parameter  $R=0.4$ is used, the jet rapidity interval is $|\eta|<1.6$, and the jet transverse momentum ranges from $45$ to $260$~GeV. So far, the only published data from the CMS collaboration that we can compare to was taken at a CM energy of $\sqrt{s}=2.76$~TeV using a jet radius parameter  $R=0.3$~\cite{Chatrchyan:2012gw}. The CMS data is presented for only one jet transverse momentum interval $100<p_T<300$~GeV and the rapidity interval is $0.3<|\eta|<2$.  Both experiments considered the distribution of light charged hadrons $h=h^+ + h^-$ inside the jet. For our numerical calculations, we use the CT14 NLO parton distribution functions~\cite{Dulat:2015mca} and the DSS07 NLO fragmentation functions~\cite{deFlorian:2007aj,deFlorian:2007hc}. In addition, we make the default scale choices of $\mu=p_T$ and $\mu_\GG=\mu_J=p_{TR}$.

In Fig.~\ref{fig:ATLAS}, we show the comparison of our calculations (solid blue lines) to the ATLAS data (red circles) taken at a CM energy of $\sqrt{s}=7$~TeV~\cite{Aad:2011sc}. The numbers in square brackets denote different jet transverse momentum bins. For example, [25, 40] means $25 < p_T < 40$ GeV. Note that the DSS07 fragmentation function are only valid for $0.05 < z_h < 1$ and $1<\mu^2< 10^5$ GeV$^2$. Beyond this range, our calculations rely on extrapolations using the distributed package from the authors~\cite{deFlorian:2007aj,deFlorian:2007hc}. Overall, we find  very good agreement between theory and data. We would like to emphasize that we do obtain improved agreement with the data in comparison to the results in~\cite{Chien:2015ctp}, where the exclusive fragmenting jet function was used to describe the same data set. While the approximations considered in~\cite{Chien:2015ctp} may not be suitable for precision phenomenology, we would like to point out that a reasonable estimate of the cross section can still be achieved. Furthermore, we find that our results here are very similar to the results of the fixed NLO calculation in~\cite{Kaufmann:2015hma}. This is to be expected, as the jet parameter $R=0.6$ is relatively large and resummation effects are less relevant.

\bef
\includegraphics[width=3.6in]{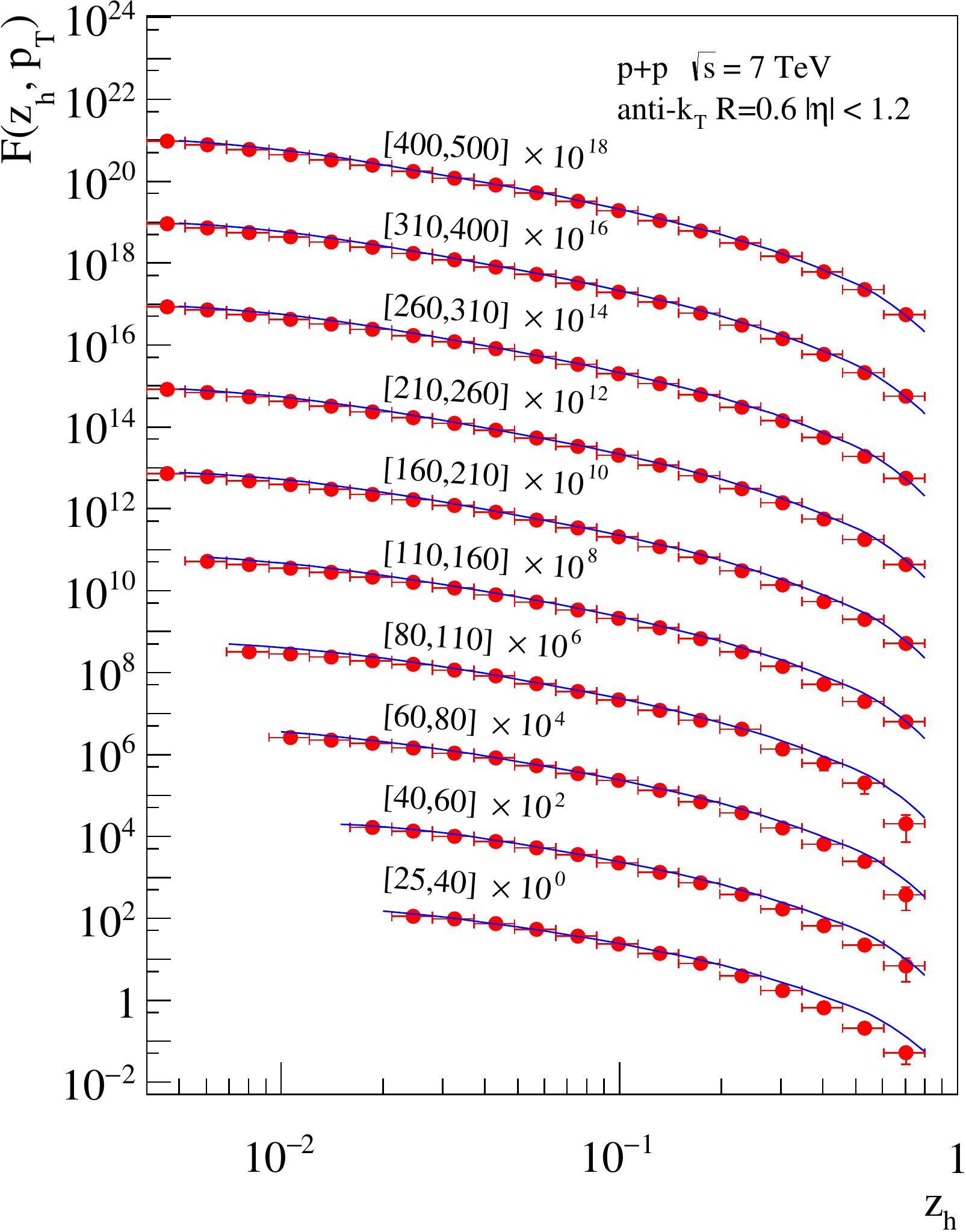}
\caption{Comparison of our numerical calculations (solid blue lines) to the ATLAS experimental data~\cite{Aad:2011sc} (red circles) in proton-proton collisions at $\sqrt{s} = 7$~TeV. Jets are reconstructed using the anti-k$_{\rm T}$~algorithm with $R=0.6$ and $|\eta|<1.2$. The numbers in the square brackets correspond to different jet transverse momentum bins in the range of $25-500$~GeV.}
\label{fig:ATLAS}
\eef

In Fig.~\ref{fig:ATLASCMS}, we compare our theoretical calculations (solid blue lines) to the preliminary ATLAS data of~\cite{ATLAS:2015mla} (red circles) and to the CMS data from Ref.~\cite{Chatrchyan:2012gw} (green triangles) in proton-proton collisions at a CM energy of $\sqrt{s} = 2.76$~TeV.  Here in particular, we find excellent agreement with the data sets for both jet radius parameters $R=0.4$ (ATLAS) and $R=0.3$ (CMS), superseding the achieved precision of the fixed NLO results in~\cite{Kaufmann:2015hma}. Especially at large-$z_h$, there is a clear indication that a fixed NLO calculation is not sufficient to describe the data. As shown below in Fig.~\ref{fig:ratio}, this is the region, where $\ln R$ resummation effects turn out to be most relevant. With these results in mind, we expect that our calculations will be very relevant for studies of the JFF in heavy-ion collisions~\cite{Chatrchyan:2012gw,Chatrchyan:2014ava,Aad:2014wha}. For these measurements, the jet radius $R$ was also chosen relatively small $R=0.2,\, 0.3$.

Note, that the CMS data shown in Fig.~\ref{fig:ATLASCMS} reaches a maximum toward low-$z_h$ and starts to fall again toward even lower $z_h$. In this region, small-$z_h$ logarithms are expected to be dominant and we can not describe the data without taking them properly into account. We are planning to address this issue in the future.

\bef
\includegraphics[width=3.6in]{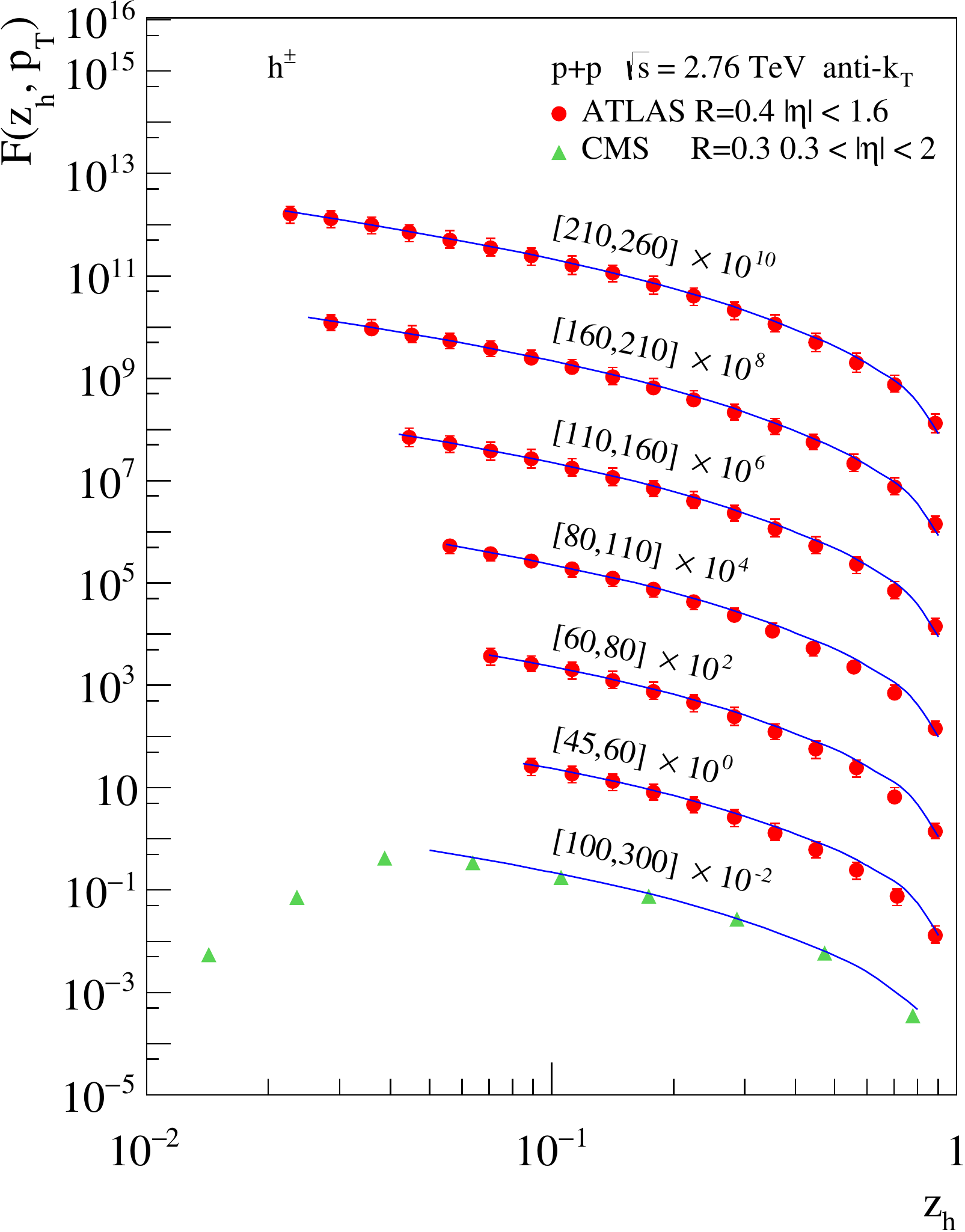}
\caption{Comparison of our numerical calculations (solid blue lines) to  LHC data in proton-proton collisions at $\sqrt{s} = 2.76$ TeV. The solid red circles correspond to the preliminary ATLAS data  form Ref.~\cite{ATLAS:2015mla} and the green triangles are the CMS data from Ref.~\cite{Chatrchyan:2012gw}.}
\label{fig:ATLASCMS}
\eef

\subsection{Comparison of NLO and NLO+NLL$_R$}

We  now  present a comparison between the fixed NLO calculation of~\cite{Kaufmann:2015hma} with our new results that include the resummation of $\ln R$. In Fig.~\ref{fig:ratio}, we show the ratio of the NLO+NLL$_R$ resummed results and the fixed NLO calculation as a function of $z_h$ for two exemplary bins of the jet transverse momentum $60<p_T<80$~GeV (left) and $260<p_T<310$~GeV (right). We choose a CM energy of $\sqrt{s}=7$~TeV, a rapidity interval of $|\eta|<1.2$ and two phenomenologically relevant values of the jet parameter $R=0.6$ (red) and $R=0.3$ (blue). In addition, we show the result for $R=0.99$ (black) illustrating that the resummed result does indeed converge to the fixed order result in the limit $R\to 1$, which is the limit of no evolution. As it turns out, the $\ln R$ resummation effects are particularly relevant for large-$z_h$ (enhancement) and small-$z_h$ (suppression). Keeping in mind that the JFF is calculated as a ratio of $\ln R$ resummed quantities, the resummation effects are in fact surprisingly large and can lead to an enhancement (or suppression) of roughly $50\%$ for $R=0.3$. The main reason is that within the SCET calculation at NLO+NLL$_R$, we evaluate the fragmentation functions at the scale $\mu_\GG=p_{T R}$. Instead, for the conventional NLO result, the FFs are evaluated at the hard scale $\mu=p_T$. Therefore, the enhancement and suppression at large-$z_h$ and small-$z_h$ respectively is consistent with a standard DGLAP-type evolution to a lower scale. As shown in Fig.~\ref{fig:ATLASCMS} above, this leads to a better agreement with the data than the fixed NLO calculation.

\bef
\includegraphics[width=2.8in]{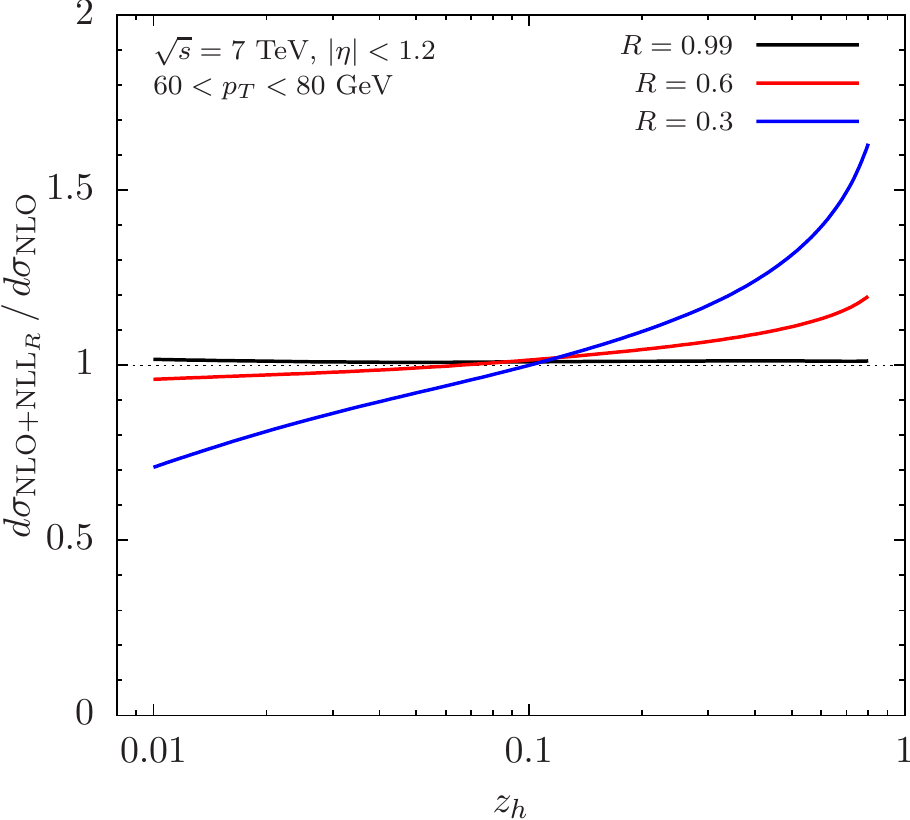} 
\hskip 0.3in
\includegraphics[width=2.8in]{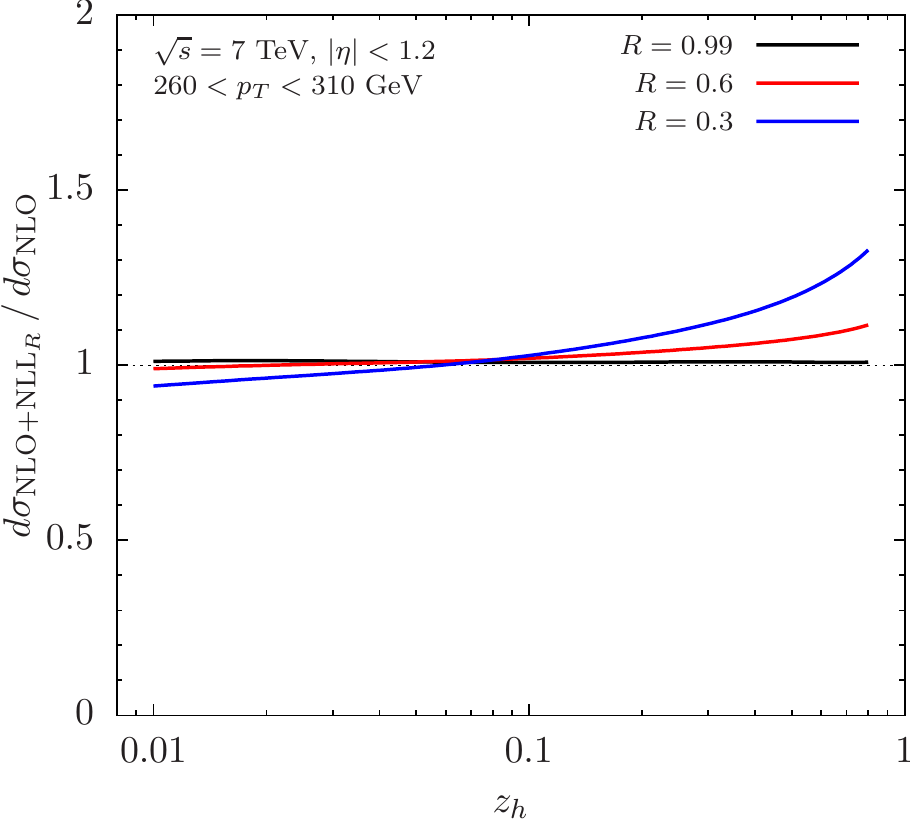} 
\caption{Ratio of the NLO+NLL$_R$ resummed cross section to the fixed NLO calculation as a function of $z_h$ for two bins of the jet transverse momentum $60<p_T<80$~GeV (left panel) and $260<p_T<310$~GeV (right panel). We choose $\sqrt{s}=7$~TeV, $|\eta|<1.2$ and two phenomenologically relevant values of the jet radius parameter $R=0.6$ (red line) and $R=0.3$ (blue line). In addition, we show the result for $R=0.99$ (black line) illustrating that our new result indeed converges to the NLO result in the limit $R\to 1$. \label{fig:ratio}}
\eef

Finally, we present results at NLO+NLL$_{R}$ accuracy for the QCD scale uncertainty in Fig.~\ref{fig:scale}. As an example, we choose the kinematics of the ATLAS data set~\cite{ATLAS:2015mla} as show in Fig.~\ref{fig:ATLASCMS}, where we have $\sqrt{s}=2.76$~TeV, $|\eta|<1.6$ and $R=0.4$. We vary all three scales $\mu,\,\mu_\GG,\,\mu_J$ independently by a factor of two around their central values $\mu=p_T$ and $\mu_{\GG,J}=p_{TR}$. We then take the envelope of these variations which is shown by the hatched red band in Fig.~\ref{fig:scale} for two sample  bins of the jet transverse momentum $60<p_T<80$~GeV (lower band) and $160<p_T<210$~GeV (upper band). All data points lie within the displayed uncertainty bands. We would like to point out an important difference compared to the results presented in~\cite{Kaufmann:2015hma,Chien:2015ctp}. At NLO and also in the resummed exclusive approximation to the JFF, there are points of vanishing scale dependence for some value of $z_h$ for any given jet transverse momentum $p_T$ and rapidity $\eta$. The vanishing scale dependence in jet cross sections is generally considered to be unphysical, see for example~\cite{Dasgupta:2014yra,Dasgupta:2016bnd}. However, in our new formalism, where single logarithms of the jet radius parameter $R$ are resummed, we do not obtain any unphysically small scale dependence.  At the same time the overall QCD scale uncertainty band does remain of similar size to the one in~\cite{Kaufmann:2015hma,Chien:2015ctp}. The absence of points with unphysical scale dependence is consistent with the results for single-inclusive jet production considered in~\cite{Dasgupta:2014yra,Dasgupta:2016bnd,Kang:2016mcy}.

\bef
\includegraphics[width=3.6in]{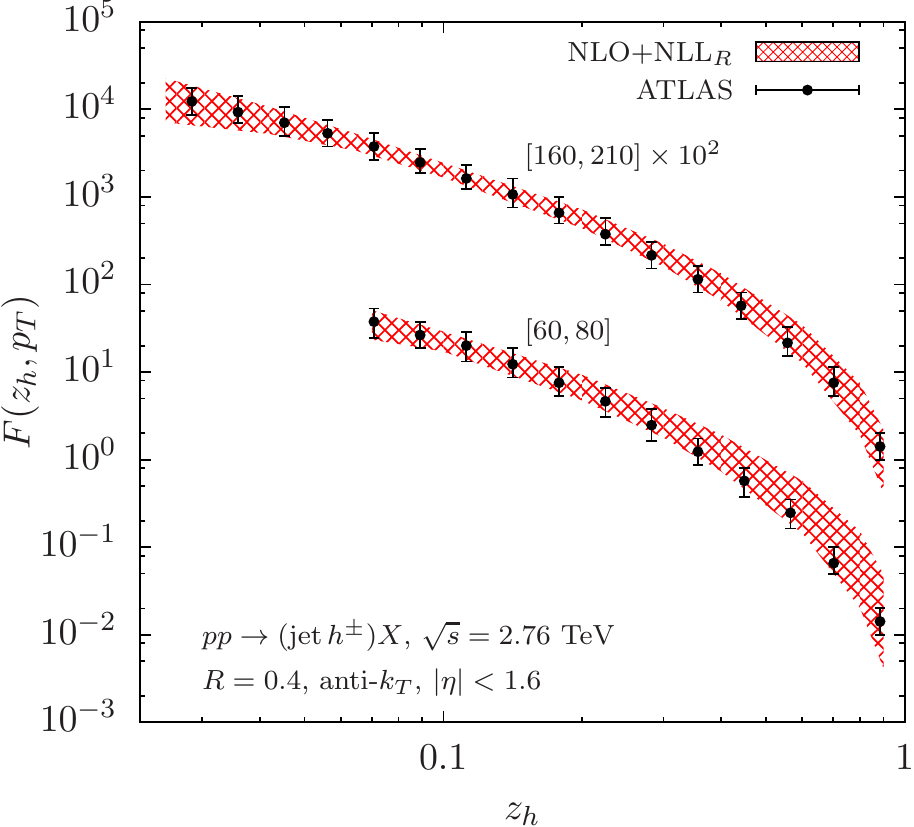} 
\caption{Results for the QCD scale uncertainty bands at NLO+NLL$_R$ accuracy using the kinematics of the ATLAS measurement~\cite{ATLAS:2015mla} as an example, as show also in Fig.~\ref{fig:ATLASCMS}. We have $\sqrt{s}=2.76$~TeV, $|\eta|<1.6$ and $R=0.4$ and we choose two sample  bins of the jet transverse momentum $60<p_T<80$~GeV (lower band) and $160<p_T<210$~GeV (upper band). We vary all three scales $\mu,\,\mu_\GG,\,\mu_J$ independently by a factor of two around their central values $\mu=p_T$ and $\mu_{\GG,J}=p_{TR}$. The envelope of these variations is shown by the hatched red band. 
\label{fig:scale}}
\eef

\section{Summary \label{sec:four}}

In this work, we introduced a new kind of jet function, the semi-inclusive fragmenting jet function. We also developed the related new formalism that allows us to calculate  the JFF as a semi-inclusive observable, rather than an exclusive one  within SCET. This approach is closer in spirit to the way in which jet substructure experimental measurements are usually performed. We found that our results without $\ln R$ resummation are consistent with previous fixed next-to-leading order (NLO) results using methods from standard perturbative QCD. We further derived DGLAP type renormalization group equations for the semi-inclusive fragmenting jet functions, which allow for a next-to-leading-logarithmic resummation of single-logarithms of the jet parameter $R$. In combination with the fixed order results, we achieved NLO+NLL$_R$ accuracy. Numerical results using this new formalism were also presented and compared to the available proton-proton data on JFFs from  LHC experiments at $\sqrt{s}=7$~TeV and $\sqrt{s}=2.76$~TeV to demonstrate excellent agreement between data and theory. Our findings are also applicable to $e^+e^-$ annihilation  and $ep$ scattering, the latter being particularly important for a future Electron-Ion Collider (EIC). In the future, we plan to extend our new formalism to other jet substructure observables, which can now be calculated as inclusive quantities rather than using exclusive approximations or through the requirement that  experiments measure exclusive $n$-jet configurations. In addition, we expect that this framework will facilitate the combination of $\ln R$ resummation with other types of resummation, such as threshold resummation. Finally, we expect significant improvements from our new framework in the ability to describe jet substructure observables in heavy-ion collisions, a rapidly growing field.

\acknowledgments
We would like to thank Tom Kaufmann, Asmita Mukherjee and Werner Vogelsang for helpful communications and for providing their NLO code for comparison. In addition, we are grateful to Christopher Lee, Yan-Qing Ma, Emanuele Mereghetti, Iain Stewart and Wouter Waalewijn for helpful discussions. This work is supported by the U.S. Department of Energy under Contract No.~DE-AC52-06NA25396, and in part by the LDRD program at Los Alamos National Laboratory. 

\bibliographystyle{JHEP}
\bibliography{bibliography}

\providecommand{\href}[2]{#2}\begingroup\raggedright\begin{thebibliography}{10}

\bibitem{Altheimer:2013yza}
A.~Altheimer et~al., {\it {Boosted objects and jet substructure at the LHC.
  Report of BOOST2012, held at IFIC Valencia, 23rd-27th of July 2012}},  {\em
  Eur. Phys. J.} {\bf C74} (2014), no.~3 2792,
  [\href{http://arxiv.org/abs/1311.2708}{{\tt arXiv:1311.2708}}].

\bibitem{Adams:2015hiv}
D.~Adams et~al., {\it {Towards an Understanding of the Correlations in Jet
  Substructure}},  {\em Eur. Phys. J.} {\bf C75} (2015), no.~9 409,
  [\href{http://arxiv.org/abs/1504.00679}{{\tt arXiv:1504.00679}}].

\bibitem{Ellis:1990ek}
S.~D. Ellis, Z.~Kunszt, and D.~E. Soper, {\it {The One Jet Inclusive
  Cross-section at Order $\alpha_s^3$ Quarks and Gluons}},  {\em Phys. Rev.
  Lett.} {\bf 64} (1990) 2121.

\bibitem{Aversa:1988vb}
F.~Aversa, P.~Chiappetta, M.~Greco, and J.~P. Guillet, {\it {QCD Corrections to
  Parton-Parton Scattering Processes}},  {\em Nucl. Phys.} {\bf B327} (1989)
  105.

\bibitem{Jager:2004jh}
B.~Jager, M.~Stratmann, and W.~Vogelsang, {\it {Single inclusive jet production
  in polarized $p p$ collisions at $O(alpha^3_s)$}},  {\em Phys. Rev.} {\bf
  D70} (2004) 034010, [\href{http://arxiv.org/abs/hep-ph/0404057}{{\tt
  hep-ph/0404057}}].

\bibitem{Mukherjee:2012uz}
A.~Mukherjee and W.~Vogelsang, {\it {Jet production in (un)polarized pp
  collisions: dependence on jet algorithm}},  {\em Phys. Rev.} {\bf D86} (2012)
  094009, [\href{http://arxiv.org/abs/1209.1785}{{\tt arXiv:1209.1785}}].

\bibitem{Currie:2013dwa}
J.~Currie, A.~Gehrmann-De~Ridder, E.~W.~N. Glover, and J.~Pires, {\it {NNLO QCD
  corrections to jet production at hadron colliders from gluon scattering}},
  {\em JHEP} {\bf 01} (2014) 110, [\href{http://arxiv.org/abs/1310.3993}{{\tt
  arXiv:1310.3993}}].

\bibitem{deFlorian:2013qia}
D.~de~Florian, P.~Hinderer, A.~Mukherjee, F.~Ringer, and W.~Vogelsang, {\it
  {Approximate next-to-next-to-leading order corrections to hadronic jet
  production}},  {\em Phys. Rev. Lett.} {\bf 112} (2014) 082001,
  [\href{http://arxiv.org/abs/1310.7192}{{\tt arXiv:1310.7192}}].

\bibitem{Dasgupta:2014yra}
M.~Dasgupta, F.~Dreyer, G.~P. Salam, and G.~Soyez, {\it {Small-radius jets to
  all orders in QCD}},  {\em JHEP} {\bf 04} (2015) 039,
  [\href{http://arxiv.org/abs/1411.5182}{{\tt arXiv:1411.5182}}].

\bibitem{Dasgupta:2016bnd}
M.~Dasgupta, F.~A. Dreyer, G.~P. Salam, and G.~Soyez, {\it {Inclusive jet
  spectrum for small-radius jets}},  {\em JHEP} {\bf 06} (2016) 057,
  [\href{http://arxiv.org/abs/1602.01110}{{\tt arXiv:1602.01110}}].

\bibitem{Kang:2016mcy}
Z.-B. Kang, F.~Ringer, and I.~Vitev, {\it {The semi-inclusive jet function in
  SCET and small radius resummation for inclusive jet production}},  {\em JHEP}
  {\bf 10} (2016) 125, [\href{http://arxiv.org/abs/1606.06732}{{\tt
  arXiv:1606.06732}}].

\bibitem{Ellis:1992qq}
S.~D. Ellis, Z.~Kunszt, and D.~E. Soper, {\it {Jets at hadron colliders at
  order $\alpha-s^{3:}$ A Look inside}},  {\em Phys. Rev. Lett.} {\bf 69}
  (1992) 3615--3618, [\href{http://arxiv.org/abs/hep-ph/9208249}{{\tt
  hep-ph/9208249}}].

\bibitem{Seymour:1997kj}
M.~H. Seymour, {\it {Jet shapes in hadron collisions: Higher orders,
  resummation and hadronization}},  {\em Nucl. Phys.} {\bf B513} (1998)
  269--300, [\href{http://arxiv.org/abs/hep-ph/9707338}{{\tt hep-ph/9707338}}].

\bibitem{Ellis:2010rwa}
S.~D. Ellis, C.~K. Vermilion, J.~R. Walsh, A.~Hornig, and C.~Lee, {\it {Jet
  Shapes and Jet Algorithms in SCET}},  {\em JHEP} {\bf 11} (2010) 101,
  [\href{http://arxiv.org/abs/1001.0014}{{\tt arXiv:1001.0014}}].

\bibitem{Procura:2009vm}
M.~Procura and I.~W. Stewart, {\it {Quark Fragmentation within an Identified
  Jet}},  {\em Phys. Rev.} {\bf D81} (2010) 074009,
  [\href{http://arxiv.org/abs/0911.4980}{{\tt arXiv:0911.4980}}]. [Erratum:
  Phys. Rev.D83,039902(2011)].

\bibitem{Liu:2010ng}
X.~Liu, {\it {SCET approach to top quark decay}},  {\em Phys. Lett.} {\bf B699}
  (2011) 87--92, [\href{http://arxiv.org/abs/1011.3872}{{\tt
  arXiv:1011.3872}}].

\bibitem{Jain:2011xz}
A.~Jain, M.~Procura, and W.~J. Waalewijn, {\it {Parton Fragmentation within an
  Identified Jet at NNLL}},  {\em JHEP} {\bf 05} (2011) 035,
  [\href{http://arxiv.org/abs/1101.4953}{{\tt arXiv:1101.4953}}].

\bibitem{Jain:2011iu}
A.~Jain, M.~Procura, and W.~J. Waalewijn, {\it {Fully-Unintegrated Parton
  Distribution and Fragmentation Functions at Perturbative $k_T$}},  {\em JHEP}
  {\bf 04} (2012) 132, [\href{http://arxiv.org/abs/1110.0839}{{\tt
  arXiv:1110.0839}}].

\bibitem{Procura:2011aq}
M.~Procura and W.~J. Waalewijn, {\it {Fragmentation in Jets: Cone and Threshold
  Effects}},  {\em Phys. Rev.} {\bf D85} (2012) 114041,
  [\href{http://arxiv.org/abs/1111.6605}{{\tt arXiv:1111.6605}}].

\bibitem{Li:2011hy}
H.-n. Li, Z.~Li, and C.~P. Yuan, {\it {QCD resummation for jet substructures}},
   {\em Phys. Rev. Lett.} {\bf 107} (2011) 152001,
  [\href{http://arxiv.org/abs/1107.4535}{{\tt arXiv:1107.4535}}].

\bibitem{Li:2012bw}
H.-n. Li, Z.~Li, and C.~P. Yuan, {\it {QCD resummation for light-particle
  jets}},  {\em Phys. Rev.} {\bf D87} (2013) 074025,
  [\href{http://arxiv.org/abs/1206.1344}{{\tt arXiv:1206.1344}}].

\bibitem{Chien:2012ur}
Y.-T. Chien, R.~Kelley, M.~D. Schwartz, and H.~X. Zhu, {\it {Resummation of Jet
  Mass at Hadron Colliders}},  {\em Phys. Rev.} {\bf D87} (2013), no.~1 014010,
  [\href{http://arxiv.org/abs/1208.0010}{{\tt arXiv:1208.0010}}].

\bibitem{Chien:2014nsa}
Y.-T. Chien and I.~Vitev, {\it {Jet Shape Resummation Using Soft-Collinear
  Effective Theory}},  {\em JHEP} {\bf 12} (2014) 061,
  [\href{http://arxiv.org/abs/1405.4293}{{\tt arXiv:1405.4293}}].

\bibitem{Hornig:2016ahz}
A.~Hornig, Y.~Makris, and T.~Mehen, {\it {Jet Shapes in Dijet Events at the LHC
  in SCET}},  {\em JHEP} {\bf 04} (2016) 097,
  [\href{http://arxiv.org/abs/1601.01319}{{\tt arXiv:1601.01319}}].

\bibitem{Bauer:2013bza}
C.~W. Bauer and E.~Mereghetti, {\it {Heavy Quark Fragmenting Jet Functions}},
  {\em JHEP} {\bf 04} (2014) 051, [\href{http://arxiv.org/abs/1312.5605}{{\tt
  arXiv:1312.5605}}].

\bibitem{Cacciari:2012mu}
M.~Cacciari, P.~Quiroga-Arias, G.~P. Salam, and G.~Soyez, {\it {Jet
  Fragmentation Function Moments in Heavy Ion Collisions}},  {\em Eur. Phys.
  J.} {\bf C73} (2013), no.~3 2319, [\href{http://arxiv.org/abs/1209.6086}{{\tt
  arXiv:1209.6086}}].

\bibitem{Ritzmann:2014mka}
M.~Ritzmann and W.~J. Waalewijn, {\it {Fragmentation in Jets at NNLO}},  {\em
  Phys. Rev.} {\bf D90} (2014), no.~5 054029,
  [\href{http://arxiv.org/abs/1407.3272}{{\tt arXiv:1407.3272}}].

\bibitem{Baumgart:2014upa}
M.~Baumgart, A.~K. Leibovich, T.~Mehen, and I.~Z. Rothstein, {\it {Probing
  Quarkonium Production Mechanisms with Jet Substructure}},  {\em JHEP} {\bf
  11} (2014) 003, [\href{http://arxiv.org/abs/1406.2295}{{\tt
  arXiv:1406.2295}}].

\bibitem{Chien:2015ctp}
Y.-T. Chien, Z.-B. Kang, F.~Ringer, I.~Vitev, and H.~Xing, {\it {Jet
  fragmentation functions in proton-proton collisions using soft-collinear
  effective theory}},  {\em JHEP} {\bf 05} (2016) 125,
  [\href{http://arxiv.org/abs/1512.06851}{{\tt arXiv:1512.06851}}].

\bibitem{Bain:2016clc}
R.~Bain, L.~Dai, A.~Hornig, A.~K. Leibovich, Y.~Makris, and T.~Mehen, {\it
  {Analytic and Monte Carlo Studies of Jets with Heavy Mesons and Quarkonia}},
  {\em JHEP} {\bf 06} (2016) 121, [\href{http://arxiv.org/abs/1603.06981}{{\tt
  arXiv:1603.06981}}].

\bibitem{Arleo:2013tya}
F.~Arleo, M.~Fontannaz, J.-P. Guillet, and C.~L. Nguyen, {\it {Probing
  fragmentation functions from same-side hadron-jet momentum correlations in
  p-p collisions}},  {\em JHEP} {\bf 04} (2014) 147,
  [\href{http://arxiv.org/abs/1311.7356}{{\tt arXiv:1311.7356}}].

\bibitem{Kaufmann:2015hma}
T.~Kaufmann, A.~Mukherjee, and W.~Vogelsang, {\it {Hadron Fragmentation Inside
  Jets in Hadronic Collisions}},  {\em Phys. Rev.} {\bf D92} (2015), no.~5
  054015, [\href{http://arxiv.org/abs/1506.01415}{{\tt arXiv:1506.01415}}].

\bibitem{Kaufmann:2016nux}
T.~Kaufmann, A.~Mukherjee, and W.~Vogelsang, {\it {Access to Photon
  Fragmentation Functions in Hadronic Jet Production}},  {\em Phys. Rev.} {\bf
  D93} (2016), no.~11 114021, [\href{http://arxiv.org/abs/1604.07175}{{\tt
  arXiv:1604.07175}}].

\bibitem{Kolodrubetz:2016dzb}
D.~W. Kolodrubetz, P.~Pietrulewicz, I.~W. Stewart, F.~J. Tackmann, and W.~J.
  Waalewijn, {\it {Factorization for Jet Radius Logarithms in Jet Mass Spectra
  at the LHC}},  \href{http://arxiv.org/abs/1605.08038}{{\tt
  arXiv:1605.08038}}.

\bibitem{Bauer:2000ew}
C.~W. Bauer, S.~Fleming, and M.~E. Luke, {\it {Summing Sudakov logarithms in $B
  \to X_s \gamma$ in effective field theory}},  {\em Phys. Rev.} {\bf D63}
  (2000) 014006, [\href{http://arxiv.org/abs/hep-ph/0005275}{{\tt
  hep-ph/0005275}}].

\bibitem{Bauer:2000yr}
C.~W. Bauer, S.~Fleming, D.~Pirjol, and I.~W. Stewart, {\it {An Effective field
  theory for collinear and soft gluons: Heavy to light decays}},  {\em Phys.
  Rev.} {\bf D63} (2001) 114020,
  [\href{http://arxiv.org/abs/hep-ph/0011336}{{\tt hep-ph/0011336}}].

\bibitem{Bauer:2001ct}
C.~W. Bauer and I.~W. Stewart, {\it {Invariant operators in collinear effective
  theory}},  {\em Phys. Lett.} {\bf B516} (2001) 134--142,
  [\href{http://arxiv.org/abs/hep-ph/0107001}{{\tt hep-ph/0107001}}].

\bibitem{Bauer:2001yt}
C.~W. Bauer, D.~Pirjol, and I.~W. Stewart, {\it {Soft collinear factorization
  in effective field theory}},  {\em Phys. Rev.} {\bf D65} (2002) 054022,
  [\href{http://arxiv.org/abs/hep-ph/0109045}{{\tt hep-ph/0109045}}].

\bibitem{Bauer:2002nz}
C.~W. Bauer, S.~Fleming, D.~Pirjol, I.~Z. Rothstein, and I.~W. Stewart, {\it
  {Hard scattering factorization from effective field theory}},  {\em Phys.
  Rev.} {\bf D66} (2002) 014017,
  [\href{http://arxiv.org/abs/hep-ph/0202088}{{\tt hep-ph/0202088}}].

\bibitem{Abazov:2008ez}
{\bf D0} Collaboration, V.~M. Abazov et~al., {\it {Measurement of differential
  $Z / \gamma^{*}$ + jet + $X$ cross sections in $p \bar{p}$ collisions at
  $\sqrt{s}$ = 1.96-TeV}},  {\em Phys. Lett.} {\bf B669} (2008) 278--286,
  [\href{http://arxiv.org/abs/0808.1296}{{\tt arXiv:0808.1296}}].

\bibitem{Group:2010aa}
{\bf CDF, D0} Collaboration, T.~T. E. V. N. P.~H. Group, {\it {Combined CDF and
  D0 Upper Limits on Standard Model Higgs-Boson Production with up to 6.7
  fb$^{-1}$ of Data}},  in {\em {Proceedings, 35th International Conference on
  High energy physics (ICHEP 2010)}}, 2010.
\newblock \href{http://arxiv.org/abs/1007.4587}{{\tt arXiv:1007.4587}}.

\bibitem{Chatrchyan:2011ne}
{\bf CMS} Collaboration, S.~Chatrchyan et~al., {\it {Jet Production Rates in
  Association with $W$ and $Z$ Bosons in $pp$ Collisions at $\sqrt{s}=7$ TeV}},
   {\em JHEP} {\bf 01} (2012) 010, [\href{http://arxiv.org/abs/1110.3226}{{\tt
  arXiv:1110.3226}}].

\bibitem{Aad:2013ysa}
{\bf ATLAS} Collaboration, G.~Aad et~al., {\it {Measurement of the production
  cross section of jets in association with a Z boson in pp collisions at
  $\sqrt{s}$ = 7 TeV with the ATLAS detector}},  {\em JHEP} {\bf 07} (2013)
  032, [\href{http://arxiv.org/abs/1304.7098}{{\tt arXiv:1304.7098}}].

\bibitem{CMS:2011ab}
{\bf CMS} Collaboration, S.~Chatrchyan et~al., {\it {Measurement of the
  Inclusive Jet Cross Section in $pp$ Collisions at $\sqrt{s}=7$ TeV}},  {\em
  Phys. Rev. Lett.} {\bf 107} (2011) 132001,
  [\href{http://arxiv.org/abs/1106.0208}{{\tt arXiv:1106.0208}}].

\bibitem{Aad:2011fc}
{\bf ATLAS} Collaboration, G.~Aad et~al., {\it {Measurement of inclusive jet
  and dijet production in $pp$ collisions at $\sqrt{s}=7$ TeV using the ATLAS
  detector}},  {\em Phys. Rev.} {\bf D86} (2012) 014022,
  [\href{http://arxiv.org/abs/1112.6297}{{\tt arXiv:1112.6297}}].

\bibitem{Chatrchyan:2012mec}
{\bf CMS} Collaboration, S.~Chatrchyan et~al., {\it {Shape, Transverse Size,
  and Charged Hadron Multiplicity of Jets in pp Collisions at 7 TeV}},  {\em
  JHEP} {\bf 06} (2012) 160, [\href{http://arxiv.org/abs/1204.3170}{{\tt
  arXiv:1204.3170}}].

\bibitem{ATLAS:2012am}
{\bf ATLAS} Collaboration, G.~Aad et~al., {\it {Jet mass and substructure of
  inclusive jets in $\sqrt{s}=7$ TeV $pp$ collisions with the ATLAS
  experiment}},  {\em JHEP} {\bf 05} (2012) 128,
  [\href{http://arxiv.org/abs/1203.4606}{{\tt arXiv:1203.4606}}].

\bibitem{ALICE:2014dla}
{\bf ALICE} Collaboration, B.~B. Abelev et~al., {\it {Charged jet cross
  sections and properties in proton-proton collisions at $\sqrt{s}=7$ TeV}},
  {\em Phys. Rev.} {\bf D91} (2015), no.~11 112012,
  [\href{http://arxiv.org/abs/1411.4969}{{\tt arXiv:1411.4969}}].

\bibitem{ATLAS:2015mla}
{\bf ATLAS} Collaboration, T.~A. collaboration, {\em {Measurement of jet
  fragmentation in 5.02 TeV proton-lead and 2.76 TeV proton-proton collisions
  with the ATLAS detector}}, 2015.
\newblock ATLAS-CONF-2015-022.

\bibitem{Aad:2011sc}
{\bf ATLAS} Collaboration, G.~Aad et~al., {\it {Measurement of the jet
  fragmentation function and transverse profile in proton-proton collisions at
  a center-of-mass energy of 7 TeV with the ATLAS detector}},  {\em Eur. Phys.
  J.} {\bf C71} (2011) 1795, [\href{http://arxiv.org/abs/1109.5816}{{\tt
  arXiv:1109.5816}}].

\bibitem{Aad:2011td}
{\bf ATLAS} Collaboration, G.~Aad et~al., {\it {Measurement of $D^{*+/-}$ meson
  production in jets from pp collisions at sqrt(s) = 7 TeV with the ATLAS
  detector}},  {\em Phys. Rev.} {\bf D85} (2012) 052005,
  [\href{http://arxiv.org/abs/1112.4432}{{\tt arXiv:1112.4432}}].

\bibitem{Chatrchyan:2012gw}
{\bf CMS} Collaboration, S.~Chatrchyan et~al., {\it {Measurement of jet
  fragmentation into charged particles in $pp$ and PbPb collisions at
  $\sqrt{s_{NN}}=2.76$ TeV}},  {\em JHEP} {\bf 10} (2012) 087,
  [\href{http://arxiv.org/abs/1205.5872}{{\tt arXiv:1205.5872}}].

\bibitem{Aad:2014wha}
{\bf ATLAS} Collaboration, G.~Aad et~al., {\it {Measurement of inclusive jet
  charged-particle fragmentation functions in Pb+Pb collisions at
  $\sqrt{s_{NN}}$=2.76 TeV with the ATLAS detector}},  {\em Phys. Lett.} {\bf
  B739} (2014) 320--342, [\href{http://arxiv.org/abs/1406.2979}{{\tt
  arXiv:1406.2979}}].

\bibitem{Chatrchyan:2014ava}
{\bf CMS} Collaboration, S.~Chatrchyan et~al., {\it {Measurement of jet
  fragmentation in PbPb and pp collisions at $\sqrt{s_{NN}}=2.76$ TeV}},  {\em
  Phys. Rev.} {\bf C90} (2014), no.~2 024908,
  [\href{http://arxiv.org/abs/1406.0932}{{\tt arXiv:1406.0932}}].

\bibitem{Abelev:2013kqa}
{\bf ALICE} Collaboration, B.~Abelev et~al., {\it {Measurement of charged jet
  suppression in Pb-Pb collisions at $\sqrt{s_{NN}}$ = 2.76 TeV}},  {\em JHEP}
  {\bf 03} (2014) 013, [\href{http://arxiv.org/abs/1311.0633}{{\tt
  arXiv:1311.0633}}].

\bibitem{Yuan:2007nd}
F.~Yuan, {\it {Azimuthal asymmetric distribution of hadrons inside a jet at
  hadron collider}},  {\em Phys. Rev. Lett.} {\bf 100} (2008) 032003,
  [\href{http://arxiv.org/abs/0709.3272}{{\tt arXiv:0709.3272}}].

\bibitem{DAlesio:2010am}
U.~D'Alesio, F.~Murgia, and C.~Pisano, {\it {Azimuthal asymmetries for hadron
  distributions inside a jet in hadronic collisions}},  {\em Phys. Rev.} {\bf
  D83} (2011) 034021, [\href{http://arxiv.org/abs/1011.2692}{{\tt
  arXiv:1011.2692}}].

\bibitem{DAlesio:2011mc}
U.~D'Alesio, L.~Gamberg, Z.-B. Kang, F.~Murgia, and C.~Pisano, {\it {Testing
  the process dependence of the Sivers function via hadron distributions inside
  a jet}},  {\em Phys. Lett.} {\bf B704} (2011) 637--640,
  [\href{http://arxiv.org/abs/1108.0827}{{\tt arXiv:1108.0827}}].

\bibitem{Aschenauer:2013woa}
E.~C. Aschenauer et~al., {\it {The RHIC Spin Program: Achievements and Future
  Opportunities}},  \href{http://arxiv.org/abs/1304.0079}{{\tt
  arXiv:1304.0079}}.

\bibitem{Aschenauer:2015eha}
E.-C. Aschenauer et~al., {\it {The RHIC SPIN Program: Achievements and Future
  Opportunities}},  \href{http://arxiv.org/abs/1501.01220}{{\tt
  arXiv:1501.01220}}.

\bibitem{Borghini:2005em}
N.~Borghini and U.~A. Wiedemann, {\it {Distorting the hump-backed plateau of
  jets with dense QCD matter}},
  \href{http://arxiv.org/abs/hep-ph/0506218}{{\tt hep-ph/0506218}}.

\bibitem{Casalderrey-Solana:2015vaa}
J.~Casalderrey-Solana, D.~C. Gulhan, J.~G. Milhano, D.~Pablos, and
  K.~Rajagopal, {\it {Predictions for Boson-Jet Observables and Fragmentation
  Function Ratios from a Hybrid Strong/Weak Coupling Model for Jet Quenching}},
   {\em JHEP} {\bf 03} (2016) 053, [\href{http://arxiv.org/abs/1508.00815}{{\tt
  arXiv:1508.00815}}].

\bibitem{medium_jet_frag}
Y.-T. Chien, Z.-B. Kang, F.~Ringer, I.~Vitev, and H.~Xing, {\em {Jet
  fragmentation function in heavy ion collisions}}, 2016.

\bibitem{He:2011pd}
Y.~He, I.~Vitev, and B.-W. Zhang, {\it {${\cal O}(\alpha_s^3)$ Analysis of
  Inclusive Jet and di-Jet Production in Heavy Ion Reactions at the Large
  Hadron Collider}},  {\em Phys. Lett.} {\bf B713} (2012) 224--232,
  [\href{http://arxiv.org/abs/1105.2566}{{\tt arXiv:1105.2566}}].

\bibitem{Chien:2015hda}
Y.-T. Chien and I.~Vitev, {\it {Towards the understanding of jet shapes and
  cross sections in heavy ion collisions using soft-collinear effective
  theory}},  {\em JHEP} {\bf 05} (2016) 023,
  [\href{http://arxiv.org/abs/1509.07257}{{\tt arXiv:1509.07257}}].

\bibitem{Chang:2016gjp}
N.-B. Chang and G.-Y. Qin, {\it {Full jet evolution in quark-gluon plasma and
  nuclear modification of jet production and jet shape in Pb+Pb collisions at
  2.76ATeV at the CERN Large Hadron Collider}},  {\em Phys. Rev.} {\bf C94}
  (2016), no.~2 024902, [\href{http://arxiv.org/abs/1603.01920}{{\tt
  arXiv:1603.01920}}].

\bibitem{Senzel:2016qau}
F.~Senzel, J.~Uphoff, Z.~Xu, and C.~Greiner, {\it {The different energy loss
  mechanisms of inclusive and b-tagged reconstructed jets within
  ultra-relativistic heavy-ion collisions}},
  \href{http://arxiv.org/abs/1602.05086}{{\tt arXiv:1602.05086}}.

\bibitem{Abe:1990ii}
{\bf CDF} Collaboration, F.~Abe et~al., {\it {Jet fragmentation properties of
  $\bar{p}p$ collisions at $\sqrt{s} = 1.8$ TeV}},  {\em Phys. Rev. Lett.} {\bf
  65} (1990) 968--971.

\bibitem{KRV}
F.~Ringer, {\em {Talk presented at 2016 QCD Evolution Workshop}}.
\newblock {Available at
  \url{https://indico.nikhef.nl/getFile.py/access?contribId=21&sessionId=1&resId=0&materialId=slides&confId=191}}.

\bibitem{Dai:2016hzf}
L.~Dai, C.~Kim, and A.~K. Leibovich, {\it {Fragmentation of a Jet with Small
  Radius}},  \href{http://arxiv.org/abs/1606.07411}{{\tt arXiv:1606.07411}}.

\bibitem{Gribov:1972ri}
V.~N. Gribov and L.~N. Lipatov, {\it {Deep inelastic e p scattering in
  perturbation theory}},  {\em Sov. J. Nucl. Phys.} {\bf 15} (1972) 438--450.
  [Yad. Fiz.15,781(1972)].

\bibitem{Lipatov:1974qm}
L.~N. Lipatov, {\it {The parton model and perturbation theory}},  {\em Sov. J.
  Nucl. Phys.} {\bf 20} (1975) 94--102. [Yad. Fiz.20,181(1974)].

\bibitem{Dokshitzer:1977sg}
Y.~L. Dokshitzer, {\it {Calculation of the Structure Functions for Deep
  Inelastic Scattering and e+ e- Annihilation by Perturbation Theory in Quantum
  Chromodynamics.}},  {\em Sov. Phys. JETP} {\bf 46} (1977) 641--653. [Zh.
  Eksp. Teor. Fiz.73,1216(1977)].

\bibitem{Altarelli:1977zs}
G.~Altarelli and G.~Parisi, {\it {Asymptotic Freedom in Parton Language}},
  {\em Nucl. Phys.} {\bf B126} (1977) 298--318.

\bibitem{Vogt:2004ns}
A.~Vogt, {\it {Efficient evolution of unpolarized and polarized parton
  distributions with QCD-PEGASUS}},  {\em Comput. Phys. Commun.} {\bf 170}
  (2005) 65--92, [\href{http://arxiv.org/abs/hep-ph/0408244}{{\tt
  hep-ph/0408244}}].

\bibitem{Anderle:2015lqa}
D.~P. Anderle, F.~Ringer, and M.~Stratmann, {\it {Fragmentation Functions at
  Next-to-Next-to-Leading Order Accuracy}},  {\em Phys. Rev.} {\bf D92} (2015),
  no.~11 114017, [\href{http://arxiv.org/abs/1510.05845}{{\tt
  arXiv:1510.05845}}].

\bibitem{Fickinger:2016rfd}
M.~Fickinger, S.~Fleming, C.~Kim, and E.~Mereghetti, {\it {Effective field
  theory approach to heavy quark fragmentation}},
  \href{http://arxiv.org/abs/1606.07737}{{\tt arXiv:1606.07737}}.

\bibitem{Jager:2002xm}
B.~Jager, A.~Schafer, M.~Stratmann, and W.~Vogelsang, {\it {Next-to-leading
  order QCD corrections to high p(T) pion production in longitudinally
  polarized pp collisions}},  {\em Phys. Rev.} {\bf D67} (2003) 054005,
  [\href{http://arxiv.org/abs/hep-ph/0211007}{{\tt hep-ph/0211007}}].

\bibitem{Collins:1989gx}
J.~C. Collins, D.~E. Soper, and G.~F. Sterman, {\it {Factorization of Hard
  Processes in QCD}},  {\em Adv. Ser. Direct. High Energy Phys.} {\bf 5} (1989)
  1--91, [\href{http://arxiv.org/abs/hep-ph/0409313}{{\tt hep-ph/0409313}}].

\bibitem{Bodwin:2015iua}
G.~T. Bodwin, K.-T. Chao, H.~S. Chung, U.-R. Kim, J.~Lee, and Y.-Q. Ma, {\it
  Fragmentation contributions to hadroproduction of prompt $j/\psi$ ,
  $\chi_{cJ}$ , and $\psi(2s)$ states},  {\em Phys. Rev.} {\bf D93} (2016),
  no.~3 034041, [\href{http://arxiv.org/abs/1509.07904}{{\tt
  arXiv:1509.07904}}].

\bibitem{Dulat:2015mca}
S.~Dulat, T.-J. Hou, J.~Gao, M.~Guzzi, J.~Huston, P.~Nadolsky, J.~Pumplin,
  C.~Schmidt, D.~Stump, and C.~P. Yuan, {\it {New parton distribution functions
  from a global analysis of quantum chromodynamics}},  {\em Phys. Rev.} {\bf
  D93} (2016), no.~3 033006, [\href{http://arxiv.org/abs/1506.07443}{{\tt
  arXiv:1506.07443}}].

\bibitem{deFlorian:2007aj}
D.~de~Florian, R.~Sassot, and M.~Stratmann, {\it {Global analysis of
  fragmentation functions for pions and kaons and their uncertainties}},  {\em
  Phys. Rev.} {\bf D75} (2007) 114010,
  [\href{http://arxiv.org/abs/hep-ph/0703242}{{\tt hep-ph/0703242}}].

\bibitem{deFlorian:2007hc}
D.~de~Florian, R.~Sassot, and M.~Stratmann, {\it {Global analysis of
  fragmentation functions for protons and charged hadrons}},  {\em Phys. Rev.}
  {\bf D76} (2007) 074033, [\href{http://arxiv.org/abs/0707.1506}{{\tt
  arXiv:0707.1506}}].

\end{thebibliography}\endgroup

\end{document}